\documentclass[aps,twocolumn,superscriptaddress,amsfont,graphicx,nofootinbib,preprintnumbers]{revtex4}
\usepackage{color,graphicx,epsfig}
\usepackage{ifpdf}
\usepackage{amsmath}
\usepackage{bm}
\usepackage{color}
\usepackage[english]{babel}
\usepackage{graphicx}
\usepackage{amsfonts}
\usepackage{amssymb}
\usepackage{hyperref}
\usepackage{enumerate}

\definecolor{nicered}{rgb}{0.7,0.1,0.1}
\definecolor{nicegreen}{rgb}{0.1,0.5,0.1}
\hypersetup{colorlinks,citecolor= nicegreen,linkcolor= nicered}

\newcommand{\beq}{\begin{equation}}
\newcommand{\eeq}{\end{equation}}
\newcommand{\bea}{\begin{eqnarray}}
\newcommand{\eea}{\end{eqnarray}}

\definecolor{Red}{rgb}{1.,0.,0.}

\def\gsim{{~\raise.15em\hbox{$>$}\kern-.85em
          \lower.35em\hbox{$\sim$}~}}
\def\lsim{{~\raise.15em\hbox{$<$}\kern-.85em
          \lower.35em\hbox{$\sim$}~}}

\newcommand{\Sc}{{\cal S}}
\newcommand{\Vc}{{\cal V}}
\newcommand{\Hc}{{\pi}}
\newcommand{\Ord}{{\mathcal O}}

\def\mysection#1{{{\bf #1}.~}}
\def\OMIT#1{}

\begin{document}

\def\Cincy{Department of Physics, University of Cincinnati, Cincinnati, Ohio 45221, USA}
\def\Perimeter{Perimeter Institute for Theoretical Physics, 31 Caroline St. N, Waterloo, Ontario, Canada N2L 2Y5}
\def\NotreDame{Department of Physics, 225 Nieuwland Science Hall, University of Notre Dame, Notre Dame, IN 46556, USA}
\def\NYU{Center for Cosmology and Particle Physics, Department of Physics, New York University, New York, NY 10003, USA}


\title{On the 750 GeV di-photon excess}

\author{Wolfgang Altmannshofer}
\email[Electronic address:]{waltmannshofer@perimeterinstitute.ca}
\affiliation{\Perimeter}

\author{Jamison Galloway}
\email[Electronic address:]{jamison.galloway@nyu.edu}
\affiliation{\NYU}

\author{Stefania  Gori}
\email[Electronic address:]{sgori@perimeterinstitute.ca}
\affiliation{\Perimeter}

\author{Alexander L. Kagan}
\email[Electronic address:]{kaganal@ucmail.uc.edu}
\affiliation{\Cincy}

\author{Adam Martin}
\email[Electronic address:]{amarti41@nd.edu}
\affiliation{\NotreDame}

\author{Jure Zupan} 
\email[Electronic address:]{zupanje@ucmail.uc.edu} 
\affiliation{\Cincy}

\begin{abstract}
We explore several perturbative scenarios in which the di-photon excess at 750 GeV can potentially be explained: a scalar singlet, a two Higgs doublet model (2HDM),  a 2HDM with an extra singlet, and  the decays of heavier resonances, both vector and scalar. We draw the following conclusions: (i) due to gauge invariance a 750 GeV scalar singlet can accommodate the observed excess more readily than a scalar $SU(2)_L$ doublet; (ii) scalar singlet production via gluon fusion is one option, however, vector boson fusion can also provide a large enough rate, 
(iii) 2HDMs with an extra singlet and no extra fermions can only give a signal in a severely tuned region of the parameter space; (iv) decays of heavier resonances can give a large enough di-photon signal at 750 GeV, while simultaneously  explaining the absence of a signal at 8 TeV. 
\end{abstract}

\maketitle
\section{Introduction} 
Recently ATLAS and CMS have observed an excess of events in the di-photon final state at 13 TeV collision energy~\cite{ATLAS-CONF-2015-081,CMS-EXO-15-004}. 
With an integrated luminosity of 3.2 ${\rm fb}^{-1}$, ATLAS has observed a $3.6\sigma$ excess at a di-photon invariant mass of $m_{\gamma\gamma}=747~{\rm GeV}$, assuming a narrow resonance~\cite{ATLAS-CONF-2015-081}. Allowing for a wide resonance, the significance increases to $3.9\sigma$ with a preferred width of approximately 45 GeV. The observed limit of the fiducial cross section at 750 GeV exceeds the expected limit by approximately 9 fb. Assuming that the resonance is a scalar produced in gluon fusion this can be interpreted as an inclusive cross section measurement of
\beq\label{eq:signalATLAS}
\sigma_{\rm ATLAS}(pp\to S\to\gamma\gamma)|_{13{\rm TeV}}=(10\pm2.8){\rm~fb}.
\eeq
CMS observed an excess using 2.6 fb${}^{-1}$ of data at 13 TeV, with a local significance of  $2.6\sigma$ peaking at $m_{\gamma\gamma}=760$ GeV, assuming a narrow width for the resonance. Assuming  a resonance with a width of 45 GeV, the local significance reduces to $2\sigma$. The corresponding cross section is
\begin{equation}
\sigma_{\rm CMS}(pp\to S\to\gamma\gamma)|_{13{\rm TeV}}=(6.5\pm3.5){\rm~fb}.
\end{equation}

 Assuming a narrow resonance, CMS also observed a roughly $2\sigma$ excess at around 750 GeV in their 8 TeV $19.7$ fb${}^{-1}$ data \cite{Khachatryan:2015qba}. The 8~TeV cross section is approximately $(0.65\pm0.35)$~fb.
ATLAS sees no significant excess in their 20.3 fb${}^{-1}$ data at 8 TeV, placing a bound at around 2.5~fb \cite{Aad:2015mna} (note that this bound is for a graviton search).

Assuming production in gluon fusion, the 8 TeV results can be translated into 13 TeV cross sections by multiplying with the ratio of gluon gluon parton luminosities at 13~TeV and 8~TeV. The corresponding cross sections at 13~TeV read 
\begin{align}
\sigma_{\rm CMS}(pp\to S\to\gamma\gamma) &= (3.1\pm1.7){\rm~fb} ~, \\
\sigma_{\rm ATLAS}(pp\to S\to\gamma\gamma) &< 11.8 {\rm~fb} ~.
\end{align}
The CMS 8~TeV result is thus in slight tension with the excess reported by ATLAS at 13~TeV, if interpreted as a resonance produced in gluon fusion.

Alternatively, assuming a production through vector boson fusion (VBF), the corresponding cross sections at 13~TeV read 

\begin{align}
\sigma_{\rm CMS}(pp\to S\to\gamma\gamma){\rm{VBF}} &= (1.6\pm 0.9){\rm~fb} ~, \\
\sigma_{\rm ATLAS}(pp\to S\to\gamma\gamma){\rm{VBF}} &< 6.3 {\rm~fb} ~,
\end{align}
showing a larger tension with  the excess reported by ATLAS at 13~TeV.
 
In this paper we discuss several new physics models that can explain the reported diphoton excess. In most of the interpretations we assume the narrow width of the resonance, but also comment on the possibility that the resonance could have a decay width of several tens of GeV.  While some aspects of our analysis may be found in the literature \cite{Falkowski:2015swt,Franceschini:2015kwy,Becirevic:2015fmu,Bellazzini:2015nxw,Gupta:2015zzs,Csaki:2015vek,Buttazzo:2015txu}, many of the results are new (for alternative interpretations see~\cite{DiChiara:2015vdm,Pilaftsis:2015ycr,Knapen:2015dap,Nakai:2015ptz,Angelescu:2015uiz,Backovic:2015fnp,Mambrini:2015wyu,Harigaya:2015ezk,Higaki:2015jag,McDermott:2015sck,Ellis:2015oso,Low:2015qep,Petersson:2015mkr,Molinaro:2015cwg,Dutta:2015wqh,Cao:2015pto,Martinez:2015kmn,Demidov:2015zqn,Chao:2015ttq,Fichet:2015vvy,Curtin:2015jcv,Bian:2015kjt,Chakrabortty:2015hff,Ahmed:2015uqt,Agrawal:2015dbf,Aloni:2015mxa,Bai:2015nbs,Cox:2015ckc,Matsuzaki:2015che,Kobakhidze:2015ldh,
No:2015bsn,Bernon:2015abk,Carpenter:2015ucu,Alves:2015jgx,Kim:2015ron,Benbrik:2015fyz,Gabrielli:2015dhk,Megias:2015ory,1410850,1410906,1410905,1410904,1410903,1410848,1410900,1410899,1410897,1410790,1410896,1410846,1410891,1410890,1410888,1410887,1410843,1410842,1410841,1410840,1410877,1410833,1410876,1410874,1410831,1410829,1410828,1410863,
1411094,1411054,1411113,1411085,1411110,1411109,1411105,1411097,Hamada:2015skp})\footnote{For earlier work see also~\cite{Jaeckel:2012yz}.}. Most importantly, we will show that Two Higgs doublet models (2HDMs) cannot accomodate the excess without introducing additional degrees of freedom. We also show that a singlet scalar produced through vector boson fusion is a viable candidate, along with the possibility already discussed in the literature -- that it is produced from gluon fusion.

The paper is organized as follows. In Sec.~\ref{sec:EFT} we discuss the observed signal in terms of an effective theory, containing a spin 0 particle with mass 750~GeV that is produced in gluon fusion and decays to diphotons through higher dimensional operators. 
In Sec.~\ref{sec:singlet} we identify the spin 0 particle with a singlet scalar that couples to gluons and photons via loops of vector-like fermions. We also comment on the possibility of production in vector boson fusion. 
In Sec.~\ref{sec:2HDM} we consider the possibility that the resonance is part of a second Higgs doublet instead of a singlet.
In Sec.~\ref{sec:BTC} we discuss scenarios where the 750~GeV resonance is produced in the decay of a more massive degree of freedom. We consider both the case of gluon fusion production of a heavy scalar and Drell-Yan production of a heavy vector. 
We conclude in Sec.~\ref{sec:conclusions}.

\section{Effective Field Theory discussion} \label{sec:EFT}

We start the discussion by forming a minimal Effective Field Theory (EFT) of the new state with the gluons and photons.
We assume CP conservation and consider the case where the new state is a scalar and comment briefly about the pseudoscalar case, since the latter leads to qualitatively similar results.
Possible interpretations of the excess in terms of a spin 2 particle are beyond the scope of this work.

Working in the phase with broken electroweak symmetry, the effective Lagrangians describing interactions of a scalar, $S$, or pseudoscalar, $A$, with gluons and photons are
\begin{eqnarray} \label{eq:LS}
{\cal L}_S &=&\lambda_g \frac{\alpha_s}{12 \pi v_W} S G_{\mu \nu}^a G^a_{\mu \nu}+\lambda_\gamma \frac{\alpha}{\pi v_W}S F_{\mu\nu} F^{\mu\nu} ~,\\ \label{eq:LA}
{\cal L}_A&=&\tilde \lambda_g \frac{\alpha_s}{12 \pi v_W} A G_{\mu \nu}^a \tilde G^a_{\mu \nu} +\tilde \lambda_\gamma \frac{\alpha}{\pi v_W}A F_{\mu\nu} \tilde F^{\mu\nu} ~,
\end{eqnarray}
where $G_{\mu\nu}^a$ and $F_{\mu\nu}$ are the gluon and photon field strengths, respectively, while $\tilde G_{\mu\nu}^a$ and $\tilde F_{\mu\nu}$ are the corresponding dual field strengths.
Note that the dimensionless couplings $\lambda_{g,\gamma}, \tilde\lambda_{g,\gamma}$ include the expected parametric loop suppressions, taking the electroweak vev, $v_W = 246$~GeV, as the generic scale.

The interactions in~(\ref{eq:LS}) allow for gluon fusion production of the scalar and lead to its decays into dijets and diphotons. 
The corresponding decay widths for the scalar read at leading order
 \begin{eqnarray}
 \Gamma_{LO}(S\to g g) &=&\frac{\alpha_S^2 }{72 \pi^3} \frac{m_S^3}{v_W^2} \lambda_g^2, \\
 \Gamma(S\to \gamma\gamma) &=&\frac{\alpha^2}{4\pi^3} \frac{m_S^3}{v_W^2} \lambda_\gamma^2,
 \end{eqnarray}
where $m_S$ is the mass of the scalar. The higher order QCD corrections to the $S \to gg$ width are large and increase $\Gamma_{LO}(S\to g g)$ by approximately 50\%~\cite{Djouadi:2005gi}.
Using $\alpha_s(m_S/2)\simeq 0.1$ and a K-factor $K\simeq 1.5$ for the gluonic decay width we obtain
 \begin{eqnarray}
 \Gamma(S\to g g) &=& 47 \text{MeV} \cdot \lambda_g^2 \cdot \Big(\frac{m_S}{750{\rm GeV}}\Big)^3 , \\
 \Gamma(S\to \gamma\gamma) &=& 3.4 \text{MeV} \cdot \lambda_\gamma^2 \cdot \Big(\frac{m_S}{750{\rm GeV}}\Big)^3 .
 \end{eqnarray}
In the limit where there are no additional decay channels with rates much larger than $S \to gg$ and $S \to \gamma \gamma$, and the theory is weakly coupled,
$S$ is a narrow resonance.

The inclusive partonic $gg\to S$ cross section is at LO given by 
 \beq
 \hat \sigma_{LO}(gg\to S)=\frac{\pi^2}{8 m_S}\Gamma_{LO}(S\to g g)\delta(\hat s -m_S^2).
 \eeq
 Using {\tt ihixs}~\cite{Anastasiou:2011pi}, we find the NNLO production cross section for $m_S=750$ GeV:
 \beq
 \sigma(pp\to S)=(590 \pm 90) {\rm fb} \cdot  \lambda_g^2,
 \eeq
where we add linearly the pdf and scale uncertainties  obtained using the 68 \% CL error estimate from {\tt MSTW2008} pdf set \cite{Martin:2009iq} and by varying $\mu\in [m_S, m_S/4]$, respectively. 

If $S\to gg$ is the dominant decay channel, the $\Gamma(S\to g g)$ cancels to first approximation in $\sigma(pp\to S) Br(S\to \gamma\gamma)$ between the production and the decay. This approximation is exact at LO, and is only approximate at higher orders due to different QCD corrections for the gluon fusion production cross section and the $S\to gg$ decay width. For the total rate into di-photon, we obtain
 \beq
 \begin{split}
 \sigma Br_{\gamma\gamma}& \simeq \sigma(pp\to S) \frac{\Gamma(S\to \gamma\gamma)}{\Gamma(S\to gg)}\\
&\simeq  590 {\rm fb}  \cdot \frac{\Gamma(S\to \gamma\gamma)}{47{\rm MeV}} \\
&\simeq  43  {\rm fb} \cdot \lambda_\gamma^2 .
 \end{split}
 \eeq

In this limit thus the diphoton excess is a measurement of the $S\to \gamma\gamma$ decay width and the EFT parameter $\lambda_\gamma$. For the ATLAS central value of $(10\pm2.8)$~fb we find
 \begin{eqnarray}
 \Gamma(S\to \gamma\gamma)& = & (0.80\pm0.25)~{\rm MeV}, \\
 \label{eq:lambda_gamma}
 \lambda_\gamma & = & 0.48\pm 0.08.
 \end{eqnarray}

\begin{figure}[t]
\centering
\includegraphics[width=0.45\textwidth]{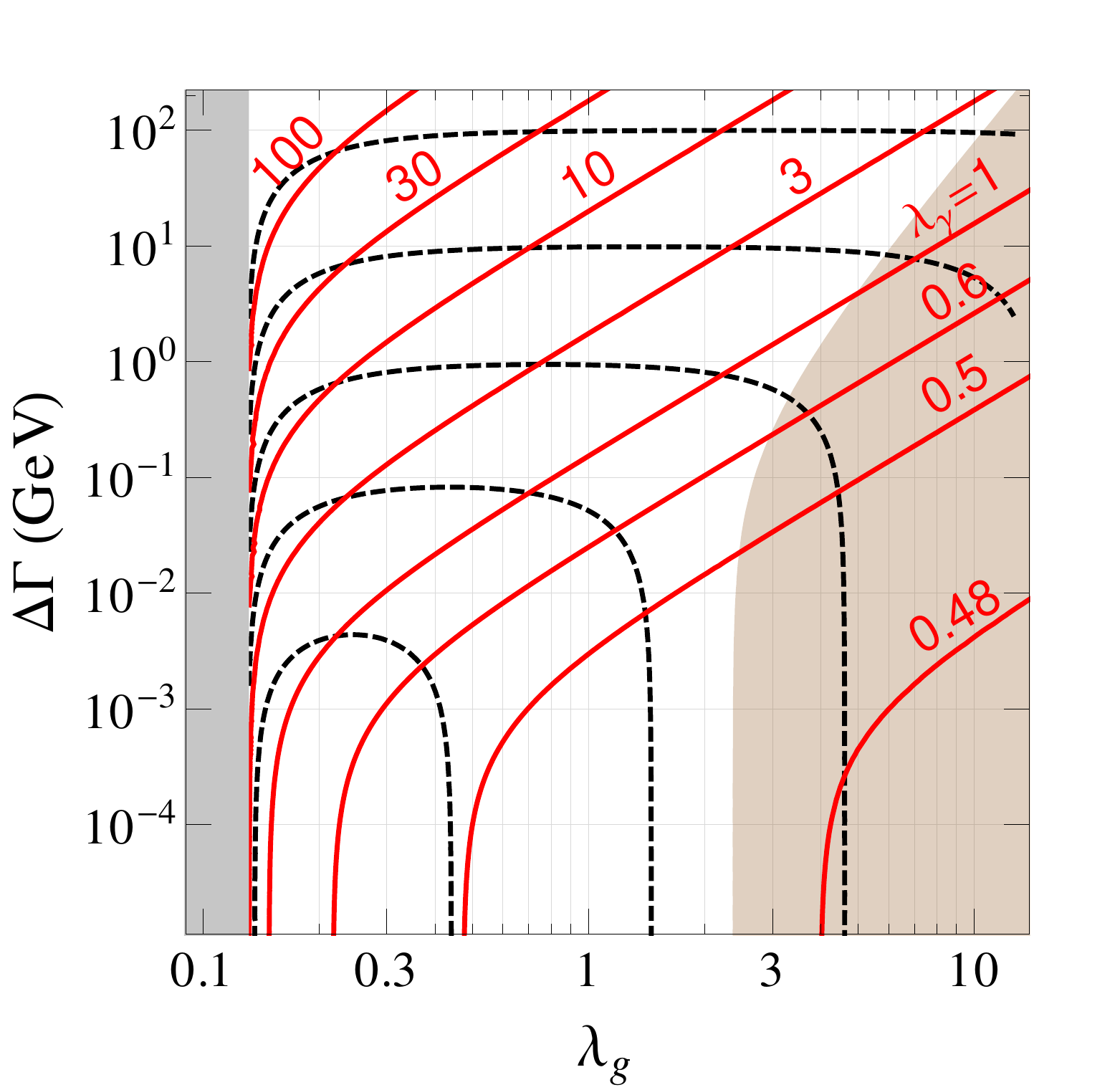}
\caption{Values of $\lambda_\gamma$ required to obtain the ATLAS central value of the observed diphoton excess, $10$~fb (red solid contours). The black dashed contours indicate the total width of the scalar $S$, from top to bottom $100/10/1/0.1/0.01$~GeV. The brown shaded region is excluded by LHC di-jet resonance searches.}
\label{fig:lambda_gamma}
\end{figure}

In general $S$ can have more interactions beyond those in~(\ref{eq:LS}), resulting in more decay channels beyond $S \to gg $ and $S \to \gamma\gamma$. Those could be into other SM particles or into NP particles. We will parameterize any additional decay width by $\Delta\Gamma$.
In Fig.~\ref{fig:lambda_gamma}, we show in the $\lambda_g$ - $\Delta\Gamma$ plane the values of $\lambda_\gamma$ required to obtain the ATLAS central value of the observed diphoton excess, $10$~fb (red solid contours), as well as the total width of the scalar $S$ (dashed black contours). In the gray region with $\lambda_g \lesssim 0.15$, the production cross section is too small to reproduce the excess, for any value of $\lambda_\gamma$. The brown shaded region produces a too large di-jet cross section to be compatible with the LHC searches of di-jet resonances~\cite{CMS-PAS-EXO-14-005}.\footnote{Note that, for this exclusion bound, we assume that the narrow width approximation describes the resonance well. At large values of $\Delta\Gamma \gtrsim 50$ GeV the actual exclusion bound will be slightly weaker than what is shown in the figure.}
For negligibly small $\Delta \Gamma$ and $\lambda_g \gtrsim 0.2$, we observe $\lambda_\gamma \simeq 0.5$ in agreement with~(\ref{eq:lambda_gamma}).
A width of $\sim 45$~GeV requires large couplings $\lambda_g, \lambda_\gamma \sim 3$. In this region of parameter space the bulk of the width comes from $\Delta \Gamma$. 
Depending on the origin of $\Delta\Gamma$, parts of the shown parameter space might be strongly constrained by direct searches for the other $S$ decay products.

\section{Scalar Singlet} \label{sec:singlet}

We now consider an explicit model that realizes the EFT described in the previous section.
We extend the SM by a single scalar singlet $S$. Due to gauge invariance, the singlet has no renormalizable couplings to the SM fermions and gauge bosons. Interactions of $S$ with the SM gauge bosons can arise from dimension 5 operators
\beq
\begin{split}\label{eq:Seffective}
{\cal L}\supset& \lambda_g \frac{\alpha_s}{12 \pi v_W} S G_{\mu \nu}^a G^a_{\mu \nu}+\lambda_B \frac{\alpha}{\pi c_W^2 v_W}S B_{\mu\nu} B^{\mu\nu}\\
&+\lambda_W \frac{\alpha}{\pi s_W^2 v_W}S W_{\mu\nu}^a W^{a\mu\nu},
\end{split}
\eeq
where $s_W=\sin\theta_W$, $c_W=\cos\theta_W$, and $\theta_W$ is the weak mixing angle.
At the renormalizable level, the singlet can have couplings to the SM Higgs doublet and mix with the Higgs after electroweak symmetry breaking \cite{Falkowski:2015swt}. This can lead to direct couplings of $S$ with $W$ and $Z$ vector bosons, as well as to a a sizable $\Delta \Gamma$ from $S\to h h$. Here, we will neglect couplings and mixing with the Higgs, and explore the consequences of dimension 5 couplings in \eqref{eq:Seffective}.

After electroweak symmetry breaking, the Lagrangian~\eqref{eq:Seffective} becomes
\beq
\begin{split}\label{eq:L:singlet:afterEWSB}
{\cal L}\supset& \lambda_g \frac{\alpha_s}{12 \pi v_W} S G_{\mu \nu}^a G^a_{\mu \nu}+\lambda_\gamma\frac{\alpha}{\pi v_W}S F_{\mu\nu} F^{\mu\nu}\\
&+\lambda_Z \frac{\alpha }{\pi  v_W} S Z_{\mu\nu} Z^{\mu\nu}+\lambda_{Z\gamma} \frac{\alpha}{\pi v_W}S Z_{\mu\nu} F^{\mu\nu}\\
&+\lambda_W \frac{2 \alpha}{\pi s_W^2 v_W}S W_{\mu\nu}^+ W^{-\mu\nu},
\end{split}
\eeq
where
\begin{align}
\lambda_\gamma&=\lambda_B+\lambda_W,\\
\lambda_{Z\gamma}&=2\Big( \lambda_W\frac{c_W}{s_W}-\lambda_B\frac{s_W}{c_W}\Big),\\
\lambda_Z&=\lambda_W\frac{c_W^2}{s_W^2}+\lambda_B\frac{s_W^2}{c_W^2}.
\end{align}
We see that generically for a nonzero $S \to \gamma\gamma$ signal one also expects the $S \to Z\gamma$, $S \to ZZ$ and $S \to W^+W^-$ decays. All these decay modes are expected to have comparable branching ratios. 

The couplings $\lambda_{g,W,B}$ can be induced from loops of additional degrees of freedom charged under the SM gauge group that couple to the singlet $S$.
For instance, we will consider the case of a vectorlike fermion with mass $m_f$, Hypercharge $Y_f$ and in $I_f$ representation of $SU(2)_L$. A coupling of this fermion to $S$ through
\beq
{\cal L}\supset -c_{f} S \bar f f,
\eeq
gives
\begin{align}
\lambda_g&= 2 c_f \frac{v_W}{m_f} C_c(r_f) D_{w}(r_f) A_f(\tau_f),\\
\lambda_B&=\frac{1}{6} c_{f} \frac{v_W}{m_f} Y_f^2 D_{w}(r_f) D_{c}(r_f) A_f(\tau_f),\\
\lambda_W&=\frac{1}{6} c_{f} \frac{v_W}{m_f} C_w(r_f)  D_{c} (r_f) A_f (\tau_f),
\end{align}
where $C_w(r_f)$ is the index of the $SU(2)_L$ representation, $Tr(T^i T^j)=C_w(r_f) \delta^{ij}$, $i,j=1,2,3$, and is $C_w(r_f)=I_f(I_f+1)D_{w}(r_f)/3$ for $D_{w}=2I_f+1$ dimensional representation of $SU(2)_L$, while $N_{c,f}$ is the dimension of the $SU(3)_c$ that the fermion belongs to, and has index $C_c(r_f)$,  $Tr(T^a T^b)=C_c(r_f) \delta^{ab}$, $a,b=1,\dots,8$. For instance, for a doublet of $SU(2)_L$ thus $C_w(2)=1/2$, $D_{w}=2$, while for a octet(triplet, singlet) of color $D_{c}=8(3,1)$, while $C_c(8)=3$, $C_c(3)=1/2$, $C_c(1)=0$. The loop function $A_f(\tau_f)$ can be found, e.g., in \cite{Carmi:2012in} and $\tau_f = m_S^2/(4 m_f^2)$.

Note that due to the dimension 5 nature of the singlet couplings to gauge bosons, the coefficients $\lambda_g$, $\lambda_B$, and $\lambda_W$ decouple as $v_W/m_f$.
In principle, any value for $\lambda_g$ and $\lambda_\gamma$ can be reproduced with a sufficiently large number of vector-like fermions. For instance, in order to reach $\lambda_g, \lambda_\gamma \sim 0.5$, the number of vector-like partners of a right-handed quark with electric charge $q$ is
\begin{eqnarray}
 N_f &\sim& 2 \cdot \left( \frac{\lambda_g}{0.5} \right) \left( \frac{1}{|c_f|} \right) \left( \frac{m_f}{1~\text{TeV}} \right) ~,\\
 N_f &\sim& 9 \cdot \left( \frac{\lambda_\gamma}{0.5} \right) \left( \frac{4}{9q^2} \right) \left( \frac{1}{|c_f|} \right) \left( \frac{m_f}{1~\text{TeV}} \right)~.
\end{eqnarray}
To reach $\lambda_\gamma \sim 0.5$, the number of vector-like partners of a right-handed lepton with charge $q$ is on the other hand
\begin{equation}
 N_f \sim 5 \cdot \left( \frac{\lambda_\gamma}{0.5} \right) \left( \frac{1}{q^2} \right) \left( \frac{1}{|c_f|} \right) \left( \frac{m_f}{500~\text{GeV}} \right) ~.
\end{equation}
This shows that, to reproduce the central value of the ATLAS excess, several new vector-like fermions at around the TeV scale would be required (or they need to have exotically large charges and/or large couplings to the singlet scalar $c_f > 1$). 

\begin{figure}[t]
\centering
\includegraphics[width=0.45\textwidth]{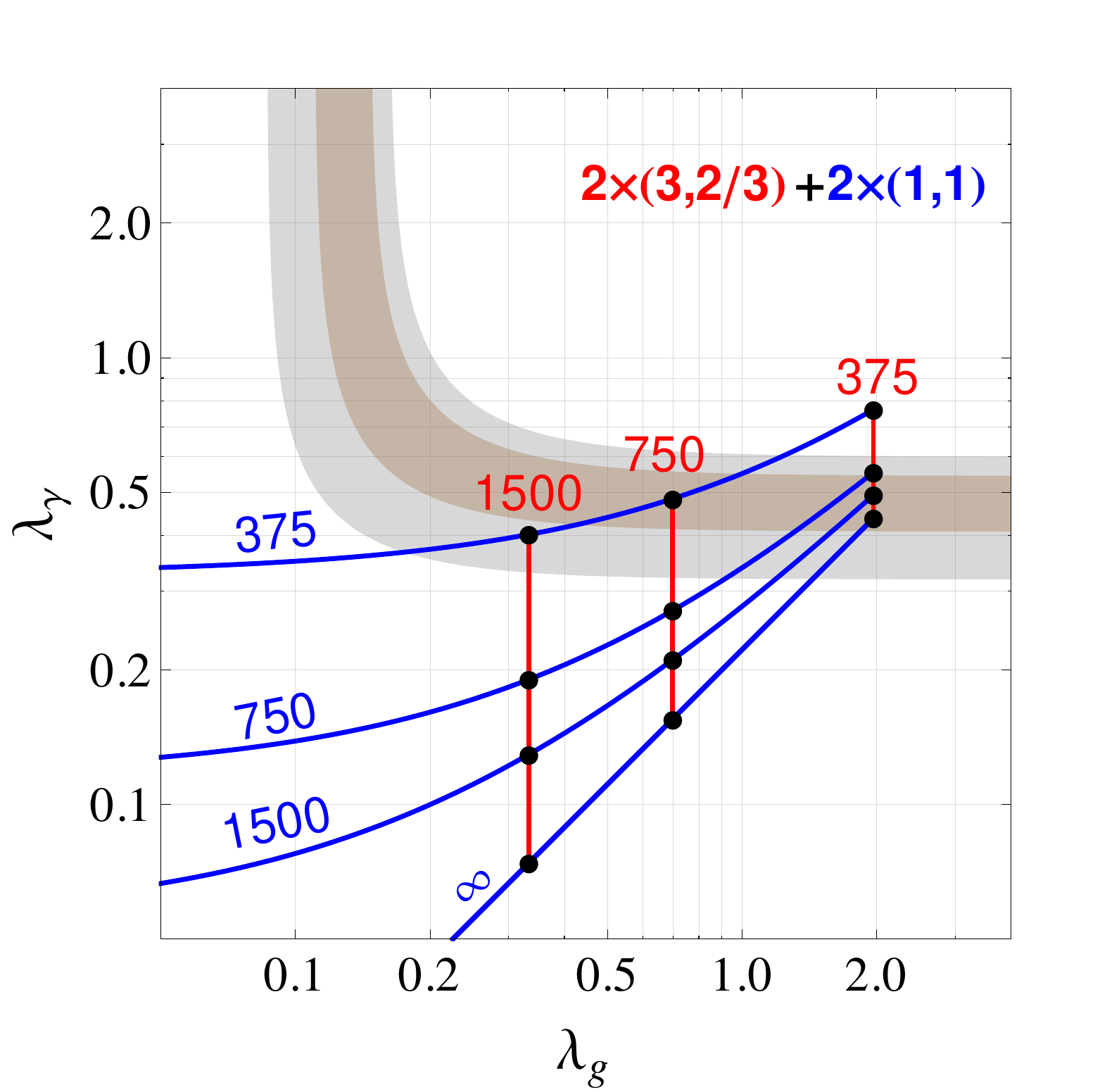} \\
\includegraphics[width=0.45\textwidth]{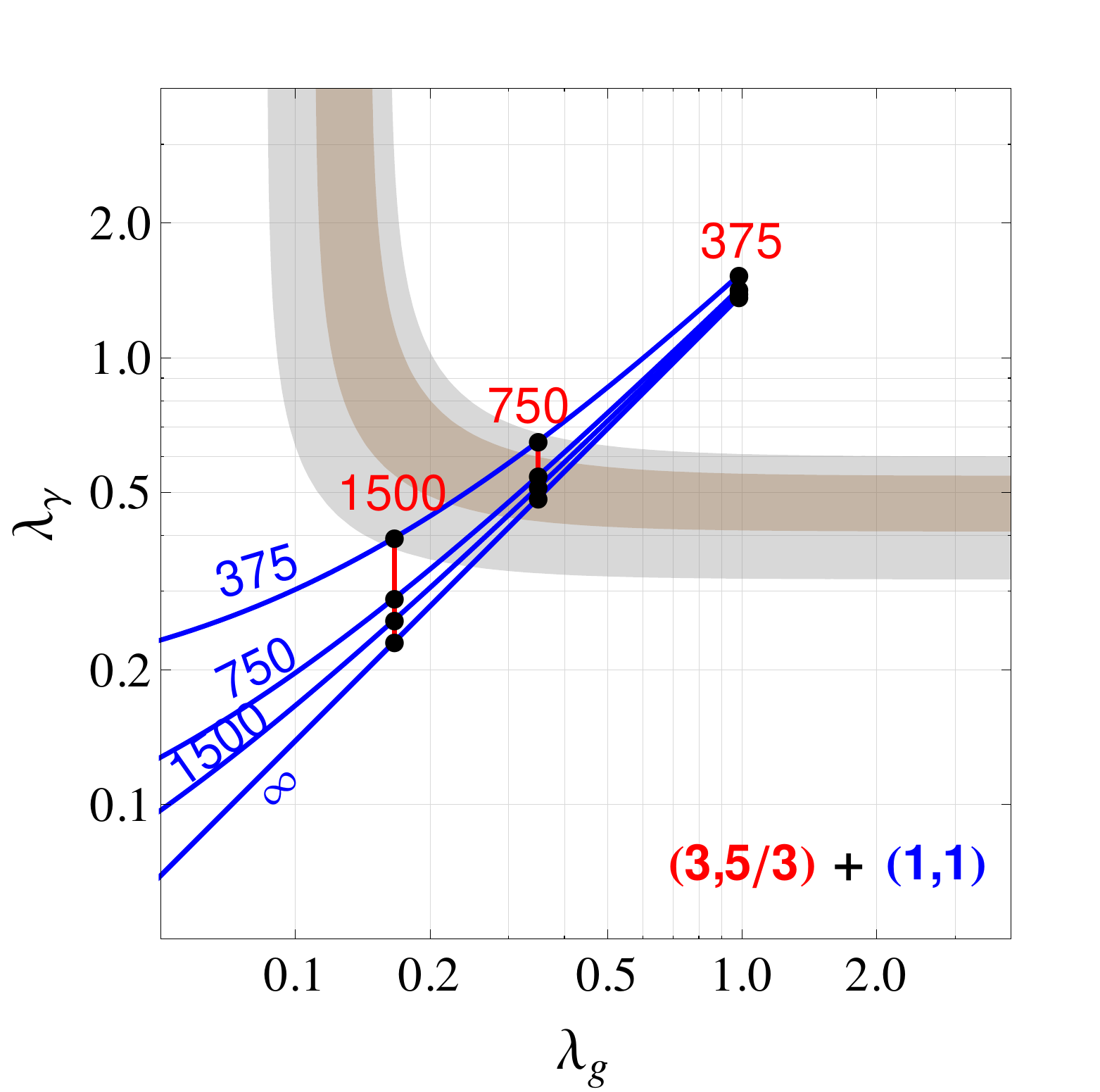}
\caption{Regions in the $\lambda_g$ - $\lambda_\gamma$ plane that can be reached in two example models with vector-like fermions. Top panel: two flavor $SU(3)$ triplets with charge $2/3$ together with two flavor $SU(3)$ singlets with charge $1$; Bottom panel: one $SU(3)$ triplet with charge $5/3$ together with one $SU(3)$ singlet with charge $1$. Setting the couplings of the fermions with the scalar $c_f=1$, the mass of the triplet/singlet in GeV is indicated by the numbers in red/blue, while red/blue lines indicate contours of fixed triplet/singlet masses. The $1\sigma$ ($2\sigma$) range of the ATLAS excess \eqref{eq:signalATLAS} gives the brown (gray) shaded region, setting $\Delta \Gamma = 0$.}
\label{fig:lambdas}
\end{figure}

In Fig.~\ref{fig:lambdas}, we present regions in the $\lambda_g$ - $\lambda_\gamma$ plane that can be reached in two example scenarios containing vector-like $SU(2)_L$ singlet fermions: 
\begin{itemize}
 \item[(i)] two flavors of color triplets with charge 2/3 together with two flavors of color singlets with charge 1 (upper plot);
 \item[(ii)] one flavor of color triplet with charge 5/3 together with one flavor of color singlet with charge 1 (lower plot).
 \end{itemize}
All fermion-scalar couplings are set to $c_f = 1$ in the plots. Blue lines represent a fixed mass of the $SU(3)$ singlet state, red lines a fixed mass of the $SU(3)$ triplet state with values indicated in GeV. (In the upper plot we assume that the two triplets and the two singlets are degenerate, for simplicity.) 
We restrict ourselves to masses above 375~GeV to kinematically forbid direct decays of the scalar $S$ into the vector-like fermions. In the shaded region in Fig.~\ref{fig:lambdas}, the diphoton excess observed by ATLAS can be reproduced, setting any additional decay width of the scalar to zero. 
Note that LHC direct searches for vector-like fermions put additional constraints on the masses of $SU(3)$ triplets. 
In the particular case of a $(3,5/3)$ representation, limits set using 2.2 fb$^{-1}$ 13 TeV data are at around 950 GeV, if the new particle decays $100\%$ into a W boson and a top quark~\cite{CMS:2015alb}. Vectorlike fermions in the $(3,2/3)$ representation, instead, have been probed up to (715-950) GeV with 8 TeV LHC data, depending on their specific decay mode ($T\to Wb,Zt,ht$)~\cite{Aad:2015kqa}. 
The precise bounds are model dependent and can be weakened if additional decay modes are present. A detailed study of the collider bounds is beyond the scope of this work.
Uncolored vectorlike fermions are much more weakly constrained. In particular, if they decay dominantly to third generation leptons, they could be as light as few (100-150)~GeV~\cite{Kumar:2015tna}.
Additional model-dependent constraints on vector-like fermions might arise from electro-weak precision observables and measurements of the properties of the 125 GeV Higgs.

\begin{figure}[t]
\centering
\includegraphics[width=0.45\textwidth]{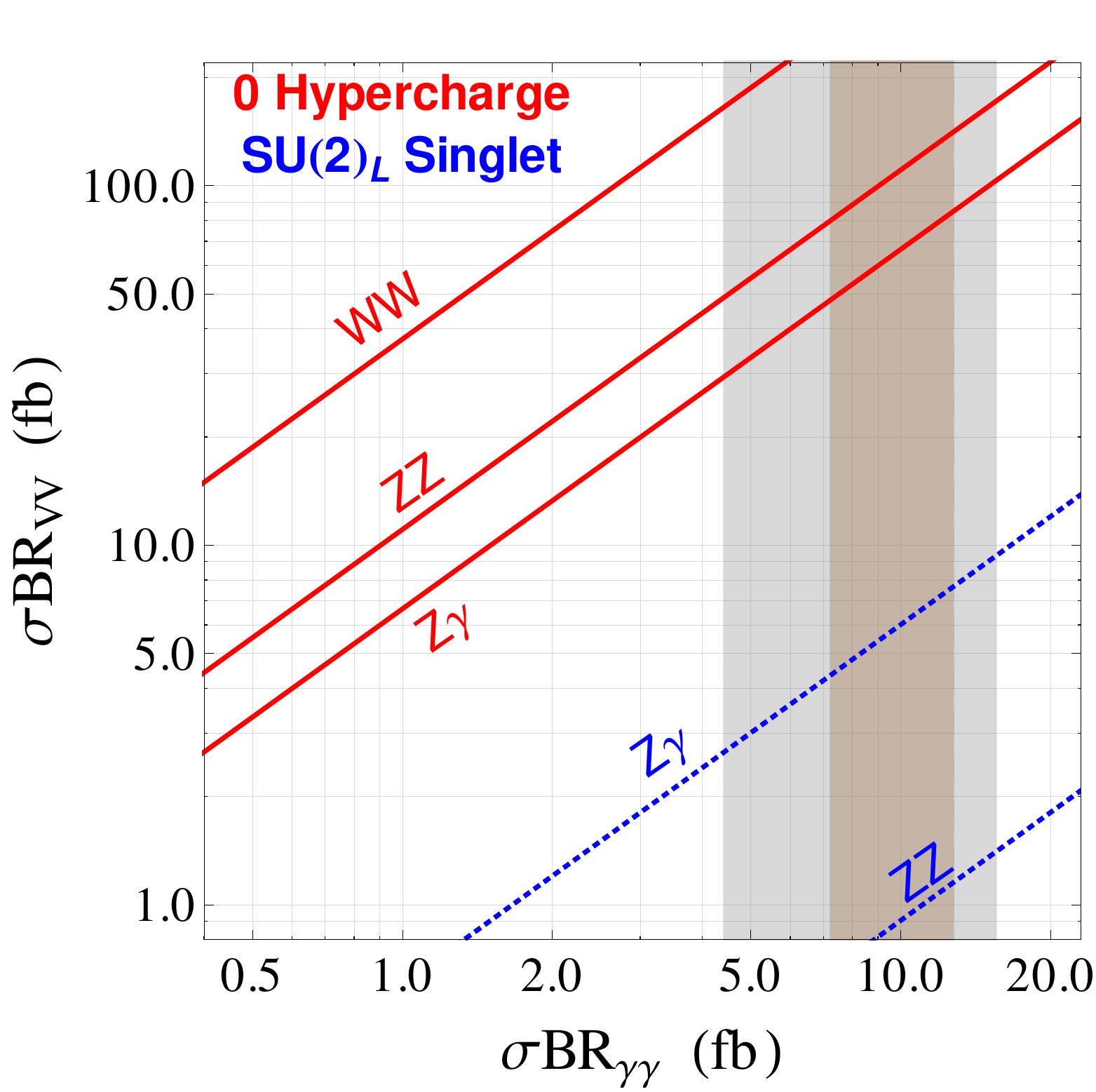}
\caption{Predicted signals for $WW$, $ZZ$, and $Z\gamma$ resonances at invariant mass of 750~GeV as function of the $\gamma\gamma$ signal in the scalar singlet model. The $1\sigma$ ($2\sigma$) range of the ATLAS excess \eqref{eq:signalATLAS} is shown in brown (gray).
Two example choices of heavy vectorlike fermions that induce the coupling to electro-weak gauge bosons are shown.
Red solid lines show the case of fermions in $SU(2)_L$ multiplets with zero Hypercharge; blue dashed lines the case of $SU(2)_L$ singlets.}
\label{fig:VVsignals}
\end{figure}

Having established that the scalar singlet allows to accommodate the observed diboson excess on condition of having several new particles charged under $SU(3)$ and $U(1)_{\rm{em}}$, we now discuss in more details the expectations for the other diboson decays $S \to Z \gamma$, $S \to ZZ$ and $S \to WW$. In terms of the couplings in \eqref{eq:L:singlet:afterEWSB} the respective branching ratios are
\beq
\begin{split}
Br_{S\to \gamma\gamma}:&Br_{S\to Z\gamma}:Br_{S\to ZZ}:Br_{S\to WW}=\\
&=2\lambda_\gamma^2:\lambda_{Z\gamma}^2:2\lambda_Z^2:  \lambda_W^2\frac{4}{s_W^4},
\end{split}
\eeq
where we neglected the small corrections due to phase space factors. Considering a simple case, where the vectorlike fermion in the loop does not carry Hypercharge, so that $\lambda_B=0$, one obtains
\beq
\begin{split}
Br_{S\to \gamma\gamma}:&Br_{S\to Z\gamma}:Br_{S\to ZZ}:Br_{S\to WW}=\\
&=1:\frac{2c_W^2}{s_W^2}:\frac{c_W^4}{s_W^4}: \frac{2}{s_W^4}\\
&\simeq 1.8\%:12\%:20\%:67\%.
\end{split}
\eeq
In this case the di-photon branching ratio is thus subleading and one would expect significant signals in the other three channels.
The corresponding cross-section times branching ratio values at 13~TeV are at the level of $\sim 100$~fb as illustrated in Fig.~\ref{fig:VVsignals}.
This case is already very constrained by existing $ZZ$ and $Z\gamma$ resonance searches~\cite{Khachatryan:2015cwa,Aad:2015kna,Aad:2014fha} that probe cross sections of $pp \to S \to ZZ$ and $pp \to S \to Z\gamma$ at the level of $\sim 15$~fb and $\sim 5$~fb at 8~TeV. At 13~TeV, this corresponds approximately to constraints of $\sim 75$~fb and $\sim 25$~fb in the $ZZ$ and $Z\gamma$ channel respectively (we are still assuming a gluon fusion production).
Also $WW$ searches~\cite{Khachatryan:2015cwa,Aad:2015agg} are already starting to be constraining. At 8~TeV, constraints on the cross section of $pp \to S \to WW$ are around 50~fb, corresponding to $\sim 250$~fb at 13~TeV. 

As another illustration, we consider the opposite limit, $\lambda_W=0$, so that the decay $S\to WW$ is absent. This is the case of the vectorlike fermions we discussed previously, that are $SU(2)_L$ singlets and thus only carry Hypercharge. Then
\beq\label{eq:gammagammalambdaB}
\begin{split}
Br_{S\to \gamma\gamma}&:Br_{S\to Z\gamma}:Br_{S\to ZZ}:Br_{S\to WW}=\\
&=1:\frac{2s_W^2}{c_W^2}:\frac{s_W^4}{c_W^4}:0\\
&\simeq 59\%:35\%:5.3\%:0\%.
\end{split}
\eeq
In this case the $\gamma\gamma$ signal is dominant, and as also illustrated in Fig.~\ref{fig:VVsignals}, the predicted $Z\gamma$ and $ZZ$ signals are at the level of few fb. 

An interesting possibility arises if none of the vectorlike fermions are charged under color. In that case $S$ does not couple to gluons, i.e., $\lambda_g=0$ in \eqref{eq:Seffective}. This possibility has been pointed in \cite{Csaki:2015vek}, where only $\lambda_\gamma$ was switched on (see also \cite{Fichet:2015vvy}). This is not possible in our framework where we keep only dimension 5 operators, since one would need to set to zero three parameters, $\lambda_{W},\lambda_Z, \lambda_{Z\gamma}$ using only two parameters, $\lambda_B$ and $\lambda_W$. The option of having $S$ couple only to photons is, however, still open if higher dimensional operators are included.

\begin{figure}[t]
\centering
\includegraphics[width=0.45\textwidth]{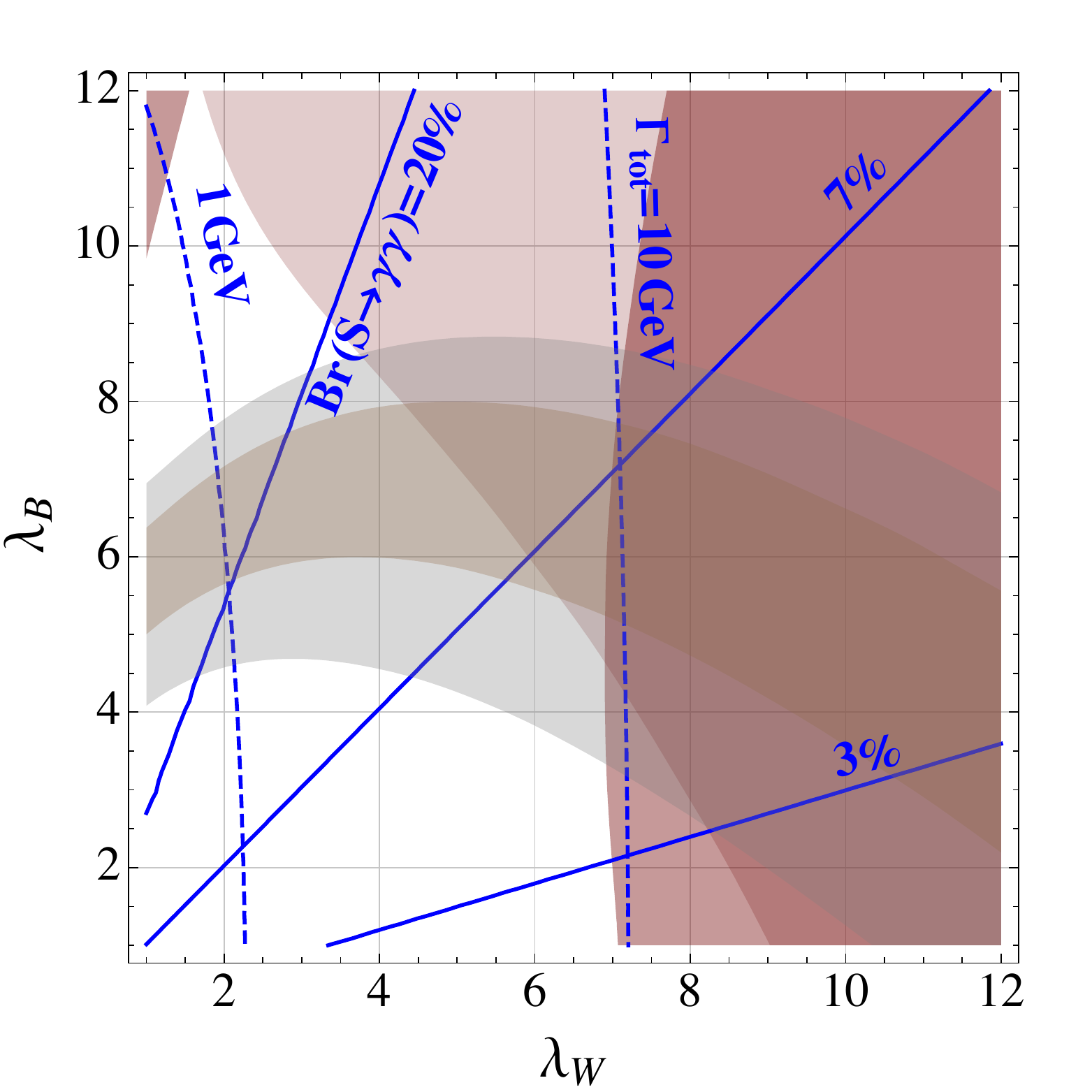}
\caption{The brown (gray) shaded region in $(\lambda_W, \lambda_B)$ parameter space that gives the $1\sigma$ ($2\sigma$) range for the diphoton signal \eqref{eq:signalATLAS} exclusively from vector boson fusion. The red shaded regions are exclusions due to 8 TeV searches in $ZZ$ \cite{Aad:2015kna,Khachatryan:2015cwa}, $Z\gamma$  \cite{Aad:2014fha} channels (lighter and darker shades, respectively).
The solid blue lines show the branching ratio to two photons, while the dashed blue lines show total decay width in GeV, assuming only $WW, ZZ, Z\gamma, \gamma\gamma$ open channels.
}
\label{fig:VBF}
\end{figure}

 In our analysis we keep only dimension 5 operators and vary $\lambda_W$ and $\lambda_B$, keeping $\lambda_g=0$. In this case $S$ can still be produced through the vector boson fusion,
$WW\to S$, $WZ\to S$, $ZZ\to S$, $Z\gamma\to S$, $W\gamma\to S$ and $\gamma\gamma\to S$. 
Working at leading order, the $WW, WZ$ and $ZZ$ fusion lead to $pp\to S j j$, the $Z\gamma$ fusion to $p p\to Sj$ and photon fusion to $pp\to S$. Adding these contributions gives the inclusive vector boson fusion cross section for $S$ production.
%
In Fig. \ref{fig:VBF} 
we show the $\lambda_B$ and $\lambda_W$ values that lead to the diphoton signal \eqref{eq:signalATLAS}. 
$\lambda_B,\lambda_W$ of order few are required. The branching ratios to two photons, $Br(S\to \gamma\gamma)=20\%,7\%,3\%$, are shown with solid blue lines (from top to bottom), while the total decay width of $S$ in GeV is given by the dashed blue lines, assuming no other open channels beyond $WW, ZZ, Z\gamma, \gamma\gamma$. When $\lambda_W\simeq \lambda_B$ the dominant production mode is photon fusion, with $WW, ZZ$ fusion a factor of two smaller, and $Z\gamma$ an order of magnitude smaller. For $\lambda_B\gg \lambda_W$ the photon fusion completely dominates, while for $\lambda_W\gg \lambda_B$ all four production modes are roughly of the same size.

Large values of $\lambda_W$ are excluded by direct searches at 8 TeV. The  regions excluded by searches in the $ZZ$ \cite{Aad:2015kna,Khachatryan:2015cwa}, and $Z\gamma$ \cite{Aad:2014fha}  channels are shaded light and dark red, respectively (the $WW$ channel searches \cite{Khachatryan:2015cwa,Aad:2015agg} lead to less stringent constraints outside the plotted parameter region). The viable parameter space in order to explain the observed diphoton  rate thus requires $\lambda_W\lesssim 7$ and $3\lesssim \lambda_B\lesssim 8$. These values are sizeable, yet still small enough that one may hope they can be realized in a concrete UV model. 

Finally, we stress that precise measurements of the various diboson rates at invariant mass of 750~GeV allow to narrow down the possible electroweak quantum numbers of the particles that mediate the couplings of the scalar $S$ to the gauge bosons. The VBF scenario in Fig. \ref{fig:VBF}  can be distinguished from the gluon fusion production by searching for the two forward jets.

\section{Two Higgs Doublet Models} \label{sec:2HDM}

We now consider the posibillity that the $750$~GeV resonance originates from a scalar $SU(2)_L$ doublet that is part of a two Higgs doublet model (2HDM). 
The two Higgs doublets contain 5 physical degrees of freedom: two neutral scalars, $h$ and $H$, one neutral pseudoscalar, $A$, and a charged Higgs, $H^\pm$, as well as the Goldstone bosons, $G^\pm$ and $G^0$, that provide longitudinal components of the $W$ and $Z$
\begin{eqnarray}
 H_1 &=& \begin{pmatrix} s_\beta G^+ - c_\beta H^+ \\ \frac{1}{\sqrt{2}} ( v_1 + c_\alpha h + s_\alpha H + i (s_\beta G^0 - c_\beta A)) \end{pmatrix}, \\
 H_2 &=& \begin{pmatrix} c_\beta G^+ + s_\beta H^+ \\ \frac{1}{\sqrt{2}} ( v_2 - s_\alpha h + c_\alpha H + i (c_\beta G^0 + s_\beta A)) \end{pmatrix}.
\end{eqnarray}
 Here $c_x = \cos x$, $s_x = \sin x$, while $t_\beta = \tan\beta = v_1 / v_2$ is the ratio of the vacuum expectation values of $H_1$ and $H_2$ with $v_W^2 = v_1^2 + v_2^2$, and $\alpha$ is the angle that diagonalizes the scalar mass matrix.
The lighter scalar $h$ is identified with the 125~GeV Higgs particle. We will discuss to which extent the heavy scalar $H$ or the pseudoscalar $A$ can re-produce the diphoton excess\footnote{We assume that CP is conserved in the scalar sector. In the presence of CP violation, $H$ and $A$ mix, which however does not qualitatively change any of our conclusions.}. Typically, one expects $H$ and $A$ to be close in mass with $m_H^2 - m_A^2 \sim v_W^2$. Therefore one might also entertain the possibility that they both contribute to the diphoton excess. With a mass splitting of few 10s of GeV, the peaks of $H$ and $A$ could even appear as one broad resonance.

We first consider the case where the second Higgs doublet is the only new degree of freedom beyond the SM, and then turn to a scenario where loops of additional degrees of freedom induce the couplings of the second doublet to photons and gluons.

\subsection{2HDM without new degrees of freedom}

In a 2HDM  where the second Higgs doublet is the only new degree of freedom beyond the SM, a possible diphoton signal is generically orders of magnitude too small compared to the observed excess. 

The coupling of $H$ and $A$ to gluons and photons are induced at the loop level by loops of SM particles. However, for a 750~GeV resonance the decays into the corresponding SM particles are kinematically allowed. This results in tiny branching ratios of $H/A \to \gamma\gamma$ at the level of $\sim 10^{-5}$. 
If we assume, for example, that the only non-negligible coupling of the pseudoscalar $A$ is to top quarks, we find
\begin{eqnarray}
 \text{BR}(A \to \gamma\gamma) &\simeq& \frac{8 \alpha_\text{em}^2}{27 \pi^2} \frac{1}{\tau_{At}} \left| f\left(\tau_{At} \right) \right|^2 \left( 1 - \frac{1}{\tau_{At}} \right)^{-1/2} \nonumber \\
 &\simeq&  8.8 \times 10^{-6} ~,
\end{eqnarray}
where $\tau_{At} = m_A^2/(4m_t^2)$ and the loop function $f$ can be found, e.g., in~\cite{Carmi:2012in}.

A very similar result holds for the scalar $H$ if its only non-negligible coupling are to top quarks
\begin{eqnarray}
 \text{BR}(H \to \gamma\gamma) &\simeq& \frac{32 \alpha_\text{em}^2}{243 \pi^2} \tau_{Ht} \left| A_f\left( \tau_{Ht} \right) \right|^2 \left( 1 - \frac{1}{\tau_{Ht}} \right)^{-3/2} \nonumber \\
 &\simeq&  8.0 \times 10^{-6} ~,
\end{eqnarray}
with $\tau_{Ht} = m_H^2/(4 m_t^2)$.

If instead, $H$ couples dominantly to weak gauge bosons we find
\begin{eqnarray}
 \text{BR}(H \to \gamma\gamma) &\simeq& \frac{49 \alpha_\text{em}^2}{16 \pi^2} \left| A_v\left( \tau_{HW} \right) \right|^2  \nonumber \\
 && \times \left(g\left( \tau_{HW} \right) + \frac{1}{2} g\left(\tau_{HZ}\right) \right)^{-1} \nonumber \\
 &\simeq&  1.8 \times 10^{-6} ~,
\end{eqnarray}
where $\tau_{HW} = m_H^2/(4 m_W^2)$, $\tau_{HZ} = m_H^2/(4 m_Z^2)$, and the kinematical function $g(1/x) = (1-x+3x^2/4) \sqrt{1-x}$. The function $A_v$ is given, e.g., in~\cite{Carmi:2012in}.

Considering sizable couplings to several fermions and/or $W$ bosons simultaneously, does not allow to increase the diphoton branching ratio appreciably. The additional loop contributions to the diphoton width are always counter-balanced by an increase in the total width due to tree-level decays into more fermions or gauge bosons. 
Note that in a 2HDM one generically also expects decays of the heavy Higgses into the 125~GeV Higgs boson, $H/A \to hh$, that will further reduce the diphoton branching ratio.

The only possibility to increase the diphoton branching ratio beyond the above estimate is through loops of charged Higgs bosons that are sufficiently heavy such that $H$ and $A$ cannot decay into $H^+ H^-$ or into $W^\pm H^\mp$.
If the charged Higgs contribution dominates, the partial decay width of $H$ into diphotons reads
\begin{equation} \label{eq:chargedhiggs}
 \Gamma(H \to \gamma\gamma) \simeq \frac{\alpha_\text{em}^2}{1024 \pi^3} \frac{v^2 m_H^3}{m_{H^\pm}^4} \left| \frac{A_s(x_{HH^\pm})}{3} \right|^2 \lambda_\pm^2 ~,
\end{equation}
where $x_{HH^\pm} = m_H^2/(4 m_{H^\pm}^2)$ and $\lambda_\pm$ is the coupling of $H$ with two charged Higgs bosons. The corresponding coupling of $A$ vanishes in the absence of CP violation. The function $A_s$ can be found for example in~\cite{Carmi:2012in}.
For $m_{H^\pm} \simeq 750$~GeV this gives $\Gamma(H \to \gamma\gamma) \simeq \lambda_\pm^2 \times 23~{\text eV} $, which is tiny. The impact of charged Higgs loops is therefore generically small even for ${\mathcal O}(1)$ couplings.
We will take into account charged Higgs loops in our numerical analysis discussed below.

The production modes of $H$ include gluon fusion, vector boson fusion, as well as production in association with gauge bosons, tops, or bottoms. The pseudoscalar $A$ can be produced in gluon fusion and in association with tops, or bottoms. For a SM-like Higgs at 750 GeV, the dominant production mode is through gluon fusion, which is in turn dominated by the top loop.  Also vector boson fusion can contribute in a non-negligible way. For the corresponding production cross-sections of $H$ we estimate
\begin{eqnarray}
 \sigma(gg\to H) &\simeq& (\xi_t^H)^2 \times \sigma(gg\to H)_\text{SM} ~, \\
 \sigma(\text{VBF}) &\simeq& (\xi_V^H)^2 \times \sigma(\text{VBF})_\text{SM} ~, \label{eq:VBF}
 \end{eqnarray}
where $\xi_t^H$, $\xi_V^H$ are the relative size of the $Htt$, $HWW$ couplings with respect to the SM top Yukawa and weak gauge coupling, respectively. The production cross-sections of a SM-like Higgs boson with mass $750$~GeV at 13 TeV proton-proton collisions are approximately $\sigma(gg \to H)_\text{SM} \simeq 620~\text{fb}$~\cite{Spira:1995mt} and $\sigma(\text{VBF})_\text{SM} \simeq 220~\text{fb}$~\cite{Dittmaier:2011ti,Dittmaier:2012vm,Heinemeyer:2013tqa}. The gluon fusion production cross section for the pseudoscalar $A$ is approximately 50\% larger~\cite{Spira:1995mt}.
Combined with a branching ratio into diphotons of at most $\Ord(10^{-5})$, this strongly suggests that a diphoton signal is orders of magnitude below the observed excess, unless the couplings of the heavy Higgses to the top quark are non-perturbatively large, $|\xi_t^H|\gg 1$. (Note that for the coupling to gauge bosons one always has $|\xi_V^H| \leq 1$ in a 2HDM.)

We performed a numerical analysis of the 2HDM parameter space, taking into account decays of $H$ and $A$ into tops, bottoms, taus, weak gauge bosons, gluons, and photons. 
In the decay to gluons we consider loops of tops and bottoms and use NLO expressions for the decay widths.
In the decay to photons we consider loops of tops, bottoms, taus, W bosons and charged Higgses. 
We considered production in gluon fusion, vector boson fusion and in association with bottom quarks. 
The gluon fusion production cross section is computed at NNLO using \verb|higlu|~\cite{Spira:1995mt} and taking into account top and bottom loops. The cross section for production in vector boson fusion is estimated as in~(\ref{eq:VBF}). To obtain the cross section for production in association with bottom quarks, we use \verb|bbh@nnlo|~\cite{Harlander:2003ai}. We neglect all other subdominant production modes. 
We scan all relevant couplings in the following generous ranges 
\begin{eqnarray}
 &&|\xi_t^{H,A}| < 1/3/5 ~,~~ |\xi_b^{H,A}| < 100 ~,~~ |\xi_\tau^{H,A}| < 200 ~, \nonumber \\
 &&|\xi_V^H|  \leq 1 ~,~~ |\lambda_\pm| < 10 ~. 
\end{eqnarray}
The reduced couplings of $H$ and $A$ to tops, bottoms and taus are taken to be independent in the scan. 
We take into account the bounds from heavy Higgs to $ZZ$ searches in~\cite{Aad:2015kna} (see also~\cite{Khachatryan:2015cwa}) that strongly constrain regions of parameter space where $|\xi_V^H|$ is sizable.
We find the following maximal signal strengths 
\begin{eqnarray}
 (\sigma\text{BR})(pp\to H\to \gamma\gamma) &\lesssim& (0.01/0.06/0.14) \,\text{fb}~, \\
 (\sigma\text{BR})(pp\to A\to \gamma\gamma) &\lesssim& (0.01/0.07/0.18) \,\text{fb}~,
\end{eqnarray}
where the first/second/third value corresponds to maximal top couplings of $|\xi_t^{H,A}| = 1/3/5$.
Adding up the $H$ and $A$ signals, we find that even with extremely large top couplings of $\xi_t^{H,A} \simeq 5$, the signal cross sections are well below 1~fb.

\subsection{Adding more degrees of freedom}

As next step, we consider additional contributions to the effective couplings of the Higgs doublets to the gauge bosons.
In an effective theory approach we write
\begin{eqnarray} \label{eq:2HDM_eff}
 \mathcal{L} &\supset& \frac{\alpha_s}{12 \pi v_W^2} \left( \lambda_1^g H_1^\dagger H_1 + \lambda_2^g H_2^\dagger H_2 \right) G_{\mu \nu}^a G^a_{\mu \nu} \nonumber \\
 && + \frac{\alpha_s}{12 \pi v_W^2} \lambda_{12}^g \left(H_1^\dagger H_2 + H_2^\dagger H_1 \right)G_{\mu \nu}^a G^a_{\mu \nu} \nonumber\\
 && - \frac{\alpha_s}{12 \pi v_W^2} \tilde \lambda_{12}^g ~i\left( H_1^\dagger H_2 - H_2^\dagger H_1 \right)G_{\mu \nu}^a \tilde G^a_{\mu \nu} ~,
\end{eqnarray}
where, as throughout the paper, we assumed that CP is conserved. 
In (\ref{eq:2HDM_eff}) we only show the couplings to gluons. Effective couplings of the Higgs doublets to the $SU(2)_L$ gauge bosons, $\lambda_{1,2,12}^W$, and to the Hypercharge gauge boson, $\lambda_{1,2,12}^B$, can be defined analogously. 
Note that the leading gauge invariant operators that couple $H_{1,2}$ to gauge bosons are of dimension 6, in contrast to the singlet case discussed above, where such couplings exists already at the dimension 5 level. 

After electroweak symmetry breaking and moving to Higgs mass eigenstates, we find for the effective couplings of $H$ and $A$ to gluons and photons
\begin{eqnarray}
 \lambda_g &=& \lambda_1^g s_\beta s_\alpha + \lambda_2^g c_\beta c_\alpha + \lambda_{12}^g (c_\beta s_\alpha + s_\beta c_\alpha)~,\\
 \lambda_\gamma &=& (\lambda_1^B + \lambda_1^W) s_\beta s_\alpha + (\lambda_2^B + \lambda_2^W) c_\beta c_\alpha \nonumber \\
 &&+ (\lambda_{12}^B + \lambda_{12}^W) (c_\beta s_\alpha + s_\beta c_\alpha)~,\\
 \tilde \lambda_g &=& \tilde \lambda_{12}^g ~,\\
 \tilde \lambda_\gamma &=& \tilde \lambda_{12}^B + \tilde \lambda_{12}^W ~.
\end{eqnarray}
In the decoupling limit, $\beta - \alpha = \pi/2$, the contributions from $\lambda_{1,2}^i$ are suppressed by $s_\beta c_\beta \to 1/t_\beta$ in the large $\tan\beta$ regime. The contributions from $\lambda_{12}^i$, on the other hand, are not suppressed at large $\tan\beta$.

As in the singlet case, the effective couplings can be induced by a multitude of new degrees of freedom.
As an example, we consider one set of vector-like quarks:
a $SU(2)_L$ doublet, $Q$,  and the corresponding singlet, $U$ (with charge $q$). The vector-like quarks couple to the second Higgs doublet $H_2$ through Yukawa interactions 
\begin{equation}
 \mathcal{L} \supset Y_2^Q H_2 \bar Q_R U_L + Y_2^U H_2 \bar Q_L U_R ~.
\end{equation}
The mass of the vector-like quarks is a sum of contributions from the $H_2$ vev, and from vector-like masses, $m_Q$ and $m_U$. For simplicity in the following we assume degenerate masses $m_Q = m_U = m$, and also set $Y_2^Q = Y_2^U = Y_2$. 
For the corresponding effective couplings to gluons and photons we find
\begin{eqnarray} \label{eq:Hgaga}
\lambda_g &=& Y_2^2 \frac{v_W^2 c_\beta c_\alpha}{m^2} A_f(\tau_{H}) ~, \\
\lambda_\gamma &=& \frac{q^2}{2} Y_2^2 \frac{v_W^2 c_\beta c_\alpha}{m^2} A_f(\tau_{H}) ~, \\
\tilde \lambda_g &=& \frac{3}{2} Y_2^2 \frac{v_W^2 c_\beta s_\beta}{m^2} \frac{1}{\tau_A} f(\tau_{A}) ~, \\\label{eq:Hgaga1}
\tilde \lambda_\gamma &=& \frac{3q^2}{4} Y_2^2 \frac{v_W^2 c_\beta s_\beta}{m^2} \frac{1}{\tau_A} f(\tau_{A}) ~.
\end{eqnarray}
The couplings of $H$ and $A$ to gluons and photons are suppressed by $1/t_\beta$ for large $\tan\beta$.
Also the couplings of the SM-like Higgs $h$ to gluons and photons are modified. However, compared to effective couplings  given above, the modifications are suppressed by $|s_\alpha/c_\alpha| \to 1/t_\beta$. Aiming for $\lambda_i, \tilde \lambda_i \sim 0.5$ and allowing for at most 10\% modifications to the couplings of $h$, implies a lower bound of roughly $\tan\beta \gtrsim 5$ in the considered setup.  

As expected, the contributions to the effective couplings in~(\ref{eq:Hgaga}) decouple with $v_W^2/m^2$ which has to be contrasted to the singlet case discussed in the previous section, where the decoupling was with $v_W/m$. Direct searches for vector-like quarks result in lower bounds on the masses of vector-like quarks at the level of 700~GeV to almost 1~TeV, depending on their decay modes~\cite{Aad:2015kqa,CMS:2015alb}. Given also the lower bound on $\tan\beta$ discussed above, we learn that a single vector-like quark which couples only to $H_2$ is by far not sufficient to accommodate the observed diphoton excess in the context of a 2HDM, unless the Yukawa coupling is non-perturbatively large $Y_2 \gg 1$. 
For $\Ord(1)$ Yukawa couplings, a very large number of vector-like quark flavors, $N_f$, would be required to reach e.g. $\lambda_g, \lambda_\gamma \sim 0.5$ 
\begin{eqnarray}
 N_f &\sim& 40 \cdot \left(\frac{\lambda_g}{0.5}\right) \left(\frac{1}{Y_2^2}\right)  \left(\frac{t_\beta}{5}\right) \left(\frac{m_f^2}{1~\text{TeV}^2}\right), \\
 N_f &\sim& 27 \cdot \left(\frac{\tilde \lambda_g}{0.5}\right) \left(\frac{1}{Y_2^2}\right)  \left(\frac{t_\beta}{5}\right) \left(\frac{m_f^2}{1~\text{TeV}^2}\right), \\
 N_f &\sim& 30 \cdot \left(\frac{\lambda_\gamma}{0.5}\right) \left(\frac{25}{9q^2}\right) \left(\frac{t_\beta}{5Y_2^2}\right)   \left(\frac{m_f^2}{1~\text{TeV}^2}\right), \\
 N_f &\sim& 20 \cdot \left(\frac{\tilde \lambda_\gamma}{0.5}\right) \left(\frac{25}{9q^2}\right) \left(\frac{t_\beta}{5 Y_2^2}\right) \left(\frac{m_f^2}{1~\text{TeV}^2}\right),
\end{eqnarray}
where we assigned the quarks a default charge of $5/3$.
Note that for such a huge number of heavy quarks, the lower bound on their mass from direct searches typically exceeds 1 TeV. 

One way to avoid the $\tan\beta$ suppression of the couplings in (\ref{eq:Hgaga})-(\ref{eq:Hgaga1}) is to couple both $H_1$ and $H_2$ to the vector-like quarks.
We add the following interactions
\begin{equation} \label{eq:Y1}
 \mathcal{L} \supset Y_1^Q H_1 \bar Q_R U_L + Y_1^U H_1 \bar Q_L U_R ~.
\end{equation}
As in the case of the Yukawa couplings of the second Higgs doublet, we will assume that both Yukawas in~(\ref{eq:Y1}) are equal $Y_1^Q = Y_1^U = Y_1$, for the sake of simplicity.

In that case we find for the loop induced couplings
\begin{eqnarray} 
\lambda_g &=& Y_1 Y_2 \frac{v_W^2 s_\beta c_\alpha}{m^2} A_f(\tau_{H}) ~, \\
\lambda_\gamma &=& \frac{q^2}{2} Y_1 Y_2 \frac{v_W^2 s_\beta c_\alpha}{m^2} A_f(\tau_{H}) ~, \\
\tilde \lambda_g &=& \frac{3}{2} Y_1 Y_2 \frac{v_W^2}{m^2} \frac{1}{\tau_A} f(\tau_{A}) ~, \\
\tilde \lambda_\gamma &=& \frac{3q^2}{4} Y_1 Y_2 \frac{v_W^2}{m^2} \frac{1}{\tau_A} f(\tau_{A}) ~.
\end{eqnarray}
At the same time also the couplings of $h$ are modified by corrections proportional to $Y_1^2$. Assuming a mild hierarchy $Y_1 < Y_2$, these corrections can be kept under control.

We give one example that allows to accommodate the diphoton excess.
For illustration, we assume that we are sufficiently close to the decoupling limit of a 2HDM, such that we can neglect the couplings of $H$ to electroweak gauge bosons. The tree-level decays to $WW$ and $ZZ$ thus do not deplete a possible $\gamma\gamma$ signal. For the coupling to the SM fermions we chose a 2HDM type I setup. In the decoupling limit the couplings of $H$ and $A$ to SM fermions are thus universally suppressed by $1/t_\beta$ compared to the SM Yukawa couplings. We also assume that couplings of $H$ and $A$ to the 125 GeV Higgs and to charged Higgs bosons are negligible. In this setup, the relevant decay modes of $H$ and $A$ are only into tops and into gluons and photons which are loop induced by the new degrees of freedom. The two scalars $H$ and $A$ are expected to be close in mass and we will assume that both contribute to the diphoton signal.

\begin{figure}[t]
\centering
\includegraphics[width=0.45\textwidth]{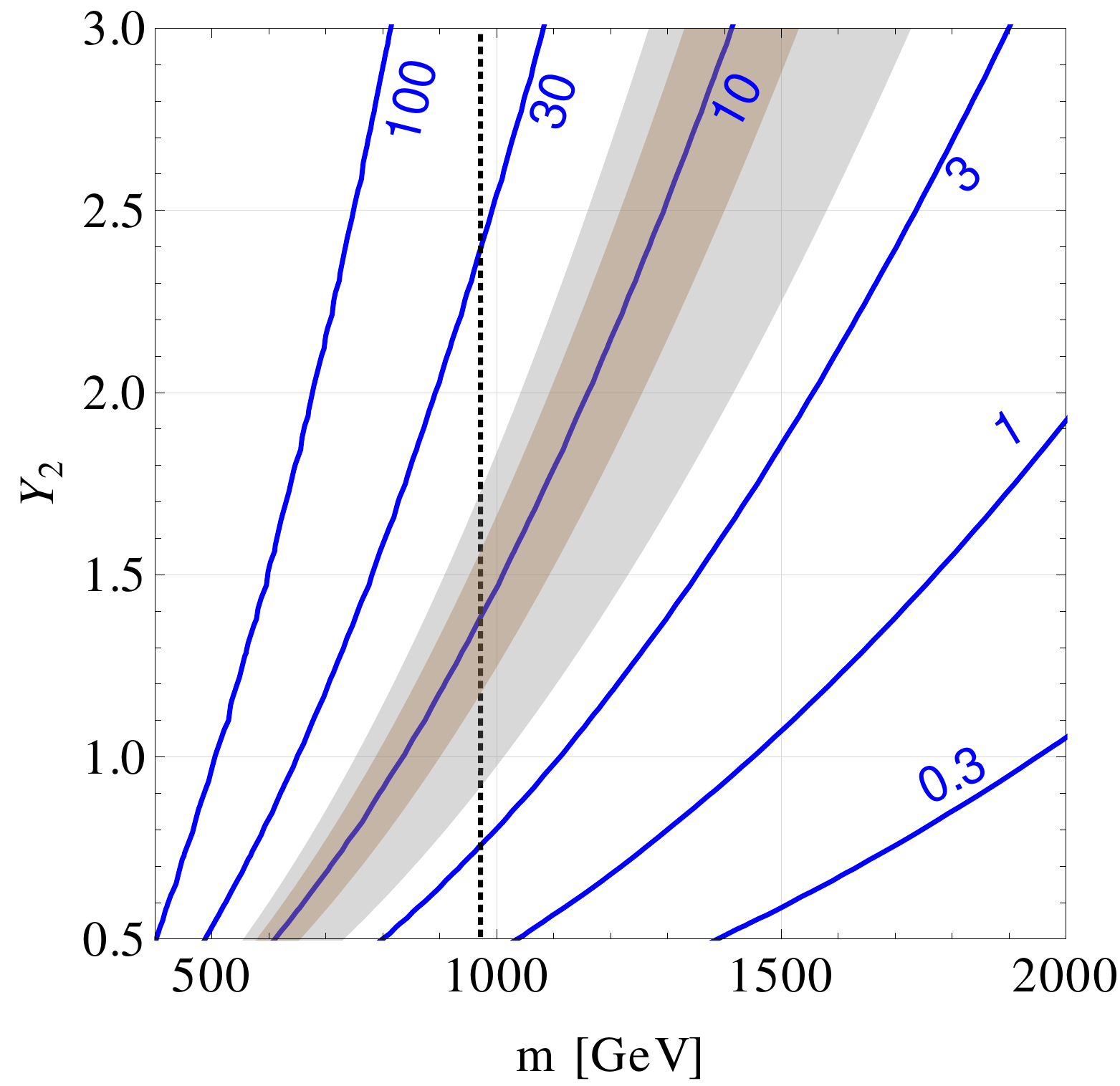}
\caption{Contours of constant cross section (in fb) for $\sigma(gg\to H)\times{\rm{BR}} (H\to\gamma\gamma) + \sigma(gg\to A)\times{\rm{BR}} (A\to\gamma\gamma)$ in the $m$ -- $Y_2$ plane for the case of 6 flavors of charge 5/3 color triplets. The value of $\tan\beta$ is 50 and the coupling $Y_1$ is set to 0.4. The shaded region corresponds to the region best fitting the ATLAS excess of events. The region on the left of the dashed black line is excluded by LHC measurements of the SM Higgs couplings to photons and gluons.}
\label{fig:2hdmplus}
\end{figure}

In Fig~\ref{fig:2hdmplus}, we show the cross section times branching ratio for the combined production of $H$ and $A$ decaying into photons in the $m$ -- $Y_2$ plane setting $Y_1 =0.4$. The photon and gluon couplings are generated by 6 flavors of vector-like charge 5/3 color triplets with degenerate mass $m$ and common Yukawa couplings $Y_{1,2}$ to the two Higgs doublets. The value of $\tan\beta$ is fixed to 50 to suppress decays of $H$ and $A$ into SM fermions, foremost tops. In the region on the left of the dashed black line the modifications to the $h$ gluon and/or photon coupling exceed 10\%.

This plot demonstrates that the observed diphoton excess could in principle come from a second Higgs doublet. However, a considerable amount of additional degrees of freedom is required to induce large enough couplings to photons and gluons. An explanation in terms of a singlet can be viable with a more minimal field content and therefore might appear more plausible.

\subsection{2HDM and a singlet scalar}

A potentially interesting possibility is that the di-photon signal is due to a cascade decay. This is possible, if one extends the 2HDM by adding a singlet scalar, $S$, without adding any additional colored or charged degrees of freedom. The signal would arise from $gg\to H\to h S(\to \gamma\gamma)$, with $m_S=750$ GeV, while $m_H> 875$ GeV. The $S\to \gamma\gamma$ is generated from $S$ coupling to the charged Higgs that runs in the loop. While in principle possible, a large enough signal is obtained only in a very tuned region of the parameter space. For instance, for $m_H=1$ TeV the production cross section is $\sigma (pp\to H)\simeq 80\, {\rm fb} \times (\xi_t^H)^2$, with $\xi_t^H$ the reduced top coupling of the heavy neutral Higgs, $H$. This means that $Br(H\to h (S\to \gamma\gamma))\sim {\mathcal O}(10\%)$ and $\xi_t^H\sim {\mathcal O}(1)$ would be required. The latter can be achieved in the small $\tan\beta$ regime of type II 2HDM, or in general type III 2HDM. A large $Br(H\to h (S\to \gamma\gamma))$ 
can be achieved 
only, if the decay widths for $H\to WW, ZZ, hh$ are negligible compared to 
$\Gamma(H\to h S)$, so that $Br(H\to h S)$ and $Br(H\to t\bar t)$ dominate. 
The $H\to WW, ZZ$ are suppressed in the alignment or decoupling
limits where $\cos(\alpha-\beta)\simeq 0$. We find that $\cos(\alpha-\beta)\sim {\mathcal O}(10^{-5})$ is required, in such a way to sufficiently suppress the $WW$ and $ZZ$ decay modes. Another requirement is that $\Gamma(S\to h h)$ is small.  This is an ad-hoc requirement, as there is no symmetry that forbids the $S hh$ coupling.  For instance, it 
arises already from the trilinear couplings $V\supset -\mu_1 S H_1^\dagger H_1 - \mu_2 S H_1^\dagger H_1  -\mu_3 S H_1^\dagger H_2 +{\rm h.c.}$. The $S hh$ coupling needs to be therefore tuned away.  The $S\to \gamma\gamma$ decay width is proportional to $\mu_i^2$. It is large enough if $\mu_{2,3}\sim 5$ TeV, with $\mu_1$ chosen such that $S\to h h $ is small. We find that tuned cancellations between $\mu_i$ at the level of ${\mathcal O}(10^{-5})$ are required. In this case $Br(S\to \gamma\gamma)\sim 7\%$. The remaining channels are $Z\gamma, ZZ, WW$, also induced through charged Higgs loops. 
In conclusion, in 2HDM with a singlet it might be possible to obtain the signal without additional fermions, albeit at the price of severe fine-tuning. Having identified corners of parameter space where a sizable diphoton signal might be possible, a detailed study would be required to ensure that additional constraints from e.g. vacuum stability, perturbativity, electro-weak precision observables, etc.  do not exclude such regions of parameter space.
If additional vector-like fermions couple to $S$ it is much easier to obtain the signal, as then $Br(S\to \gamma\gamma)$ can be $\sim{\mathcal O}(1)$.

The situation is even more dire in three Higgs doublet models. In principle the light component of an additional Higgs doublet could play a similar role as the singlet above, leading to $pp \to H\to h h_3\to \gamma\gamma$. However, the situation is even more challenging, as one cannot obtain large enough trilinear $h_3 H^+H^-$ couplings without violating unitarity constraints. We thus found no viable solutions for the diphoton excess in this case.
\section{The signal from higher mass resonances} \label{sec:BTC}

So far we mainly discussed di-photon signals from direct resonant production $pp\to X \to \gamma\gamma$. Cascade decays, of the form $pp\to X\to Y(\to\gamma\gamma) Y'$, have several beneficial features compared to direct resonant production. Since the production of $X$ and the decay of $Y$ are in principle unrelated it is easy to achieve large $Br(Y\to \gamma\gamma)$.
In addition, if $X$ is heavy enough, this can explain the slight tension between 8 and 13 TeV data, due to the absence of a sizable excess at 8 TeV. There are two distinct production mechanisms that we consider, (i) production through gluon fusion and (ii) the Drell-Yan production. Each of these has a significantly increased cross section when going from 8 TeV to 13 TeV. For gluon fusion the ratio of parton luminosities is 5.9 (15.0) for $m_X=1\,{\rm TeV}\, (2\, {\rm TeV})$, while for the Drell-Yan production the corresponding ratio of $q\bar q$ parton luminosities is 3.2 (7.6). For 2 TeV $X$ resonance produced from gluon fusion the sensitivity of the present 13 TeV diphoton searches is thus larger 
than the 8 TeV diphoton searches. Note that any cascade decays require that there are other objects in the event beside the photons. The status of whether or not there are other objects in the event is unclear at the moment, though in future they could rule in or out the cascade explanation of the di-photon excess.

\subsection{Gluon fusion production}
We first discuss a phenomenological model where the production is dominated by gluon fusion. The model consists of two scalars, $S_1$ and $S_2$, taken to be SM gauge singlets. The di-photon signal is produced from the $pp\to S_2\to S_1(\to \gamma\gamma) S_1$ decay chain, where $S_1$ has a mass of $m_{S_1}=750$ GeV.  The interaction Lagrangian is assumed to contain the coupling
\beq\label{eq:S2S1}
{\cal L}_{\rm int}\supset \lambda_S m_{S_2} S_2 S_1^2,
\eeq
which leads to the $S_2\to S_1 S_1$ decay. In this toy model we assume that $S_1$ and $S_2$ couple to SM only through higher dimensional operators containing the gauge fields, \eqref{eq:Seffective},\eqref{eq:L:singlet:afterEWSB}, with the obvious generalization of the notation, $\lambda_B\to\lambda_{i,B}$,..., with $i=1,2$. For natural values of the interaction \eqref{eq:S2S1}, $\lambda_S\sim {\mathcal O}(1)$, one has $Br(S_2\to S_1 S_1)\simeq 100\%$ for $m_{S_2}\gtrsim 2 m_{S_1}=1.5{\rm~TeV}$.

\begin{figure}[t]
\centering
\includegraphics[width=0.45\textwidth]{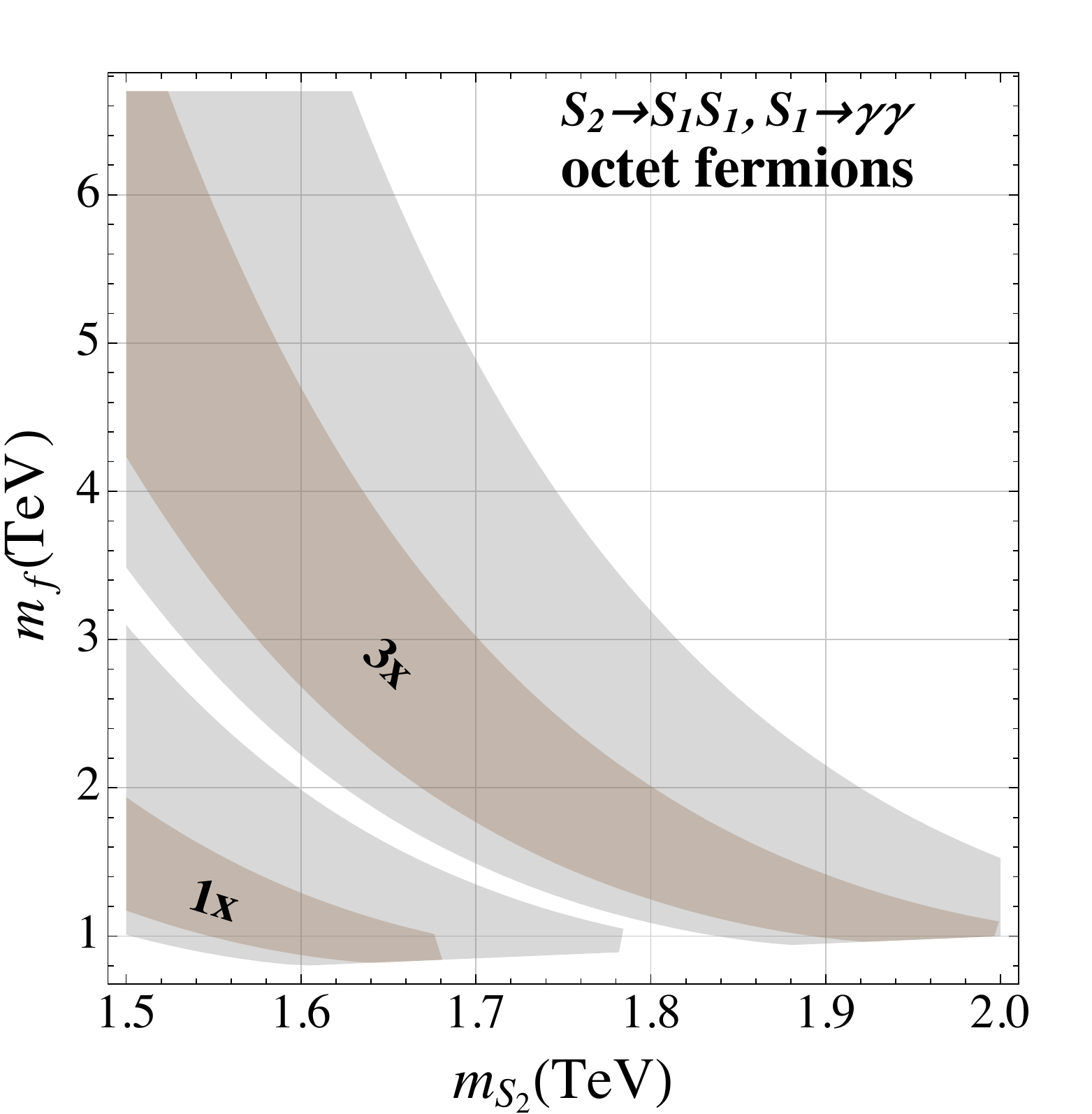}
\caption{The mass of $N_f=1,3$ copies copies  of  color octet vector-like fermions coupling to $S_2$ resonance with $g_{2,f}=1$ strength, such that a di-photon signal \eqref{eq:signalATLAS} is obtained from $pp\to S_2\to S_1 S_1$ decay chain, where $S_1$ decays to photons through loops of fermions that carry only Hypercharge.  }
\label{fig:mS2mf}
\end{figure}

In this example it is easy to obtain a large enough di-photon signal. In Fig. \ref{fig:mS2mf} we show the scenario where $S_2$ couples to gluons, while $S_1$ only couples to $B_\mu$. That is, the effective Lagrangian is
\beq
{\cal L}_{\rm eff}= \lambda_{2,g} \frac{\alpha_s}{12 \pi v_W} S_2 G_{\mu\nu}^a G_{\mu\nu}^a +\lambda_{1,B} \frac{\alpha}{\pi c_W^2 v_W} S_1 B_{\mu\nu} B_{\mu\nu},
\eeq
while the remaining dimension five operators are set to zero for simplicity. Renormalizable realizations of this Lagrangian are models in which the $S_2$ couples to $N_f$ vector-like fermions charged under $SU(3)_c$, 
\beq
{\cal L}_{2}=- \sum_{i=1}^{N_f} g_{2,f}^i S_2 \bar f_i f_i,
\eeq
which gives
\beq
\lambda_{2,g}= 2 C_c(r_f) \frac{g_{2,f} v_W}{m_f} A_f(\tau_f),
\eeq
while $S_1$ couples to a different set of vector-like fermions that only carry Hypercharge. This means that the branching ratio to photons is $Br(S_1\to \gamma\gamma)=59\%$, cf. Eq. \eqref{eq:gammagammalambdaB}, irrespective of the Hypercharges of the fermions. 

In Fig. \ref{fig:mS2mf} we set $g_{2,f}^i=1$, and take the colored vector-like fermions to be octets of $SU(3)_c$ and have the same mass, $m_f$. We then show the dependence of $m_f$ on the mass of the first resonance in the decay chain, $m_{S_2}$, for $N_f=1,3$ generations of fermions (from bottom to top). The brown (gray) bands correspond to a $1\sigma$ $(2\sigma)$ range of the diphoton signal \eqref{eq:signalATLAS}. We see that for a $\sim 2$ TeV resonance, $S_2$, three generations of fermions are required with mass $\sim 1$TeV (or a correspondingly higher coupling constant $g_{2,f}\sim 3$). For other charge assignments of the vector-like fermions the details of the numerics change. If the fermions coupling to $S_1$ carry only $SU(2)_L$ charges, 
then the branching ratio is much smaller, $Br(S_1\to \gamma\gamma)=1.8\%$, which would need to be compensated by a significantly higher gluonic cross section either due to larger $g_{2,f}$, or by having more generations, or both.

Finally, we comment on the possibility that the diphoton excess is coming from very light $S_1$ (see also a discussion in \cite{Knapen:2015dap}). If $m_{S_1}\ll m_{S_2}$ the two photons can merge and lead to an effective diphoton signal in the detector. Taking granularity of the electromagnetic calorimeter as the guidance, the two photons merge for $m_{S_2}=750$ GeV, if $m_{S_1}\lesssim 20$ GeV. The photons can still be resolved, however, if both photons convert in the inner tracker. The probability of this to happen is ${\mathcal O}(20\%)$. A search for the $m_{S_1}$ peak in the $m_{\gamma\gamma}$ distribution using only converted photons can thus reveal this possibility with about five times the present statistics.  

\subsection{Drell-Yan production}

\begin{figure}[t]
\centering
\includegraphics[width=0.45\textwidth]{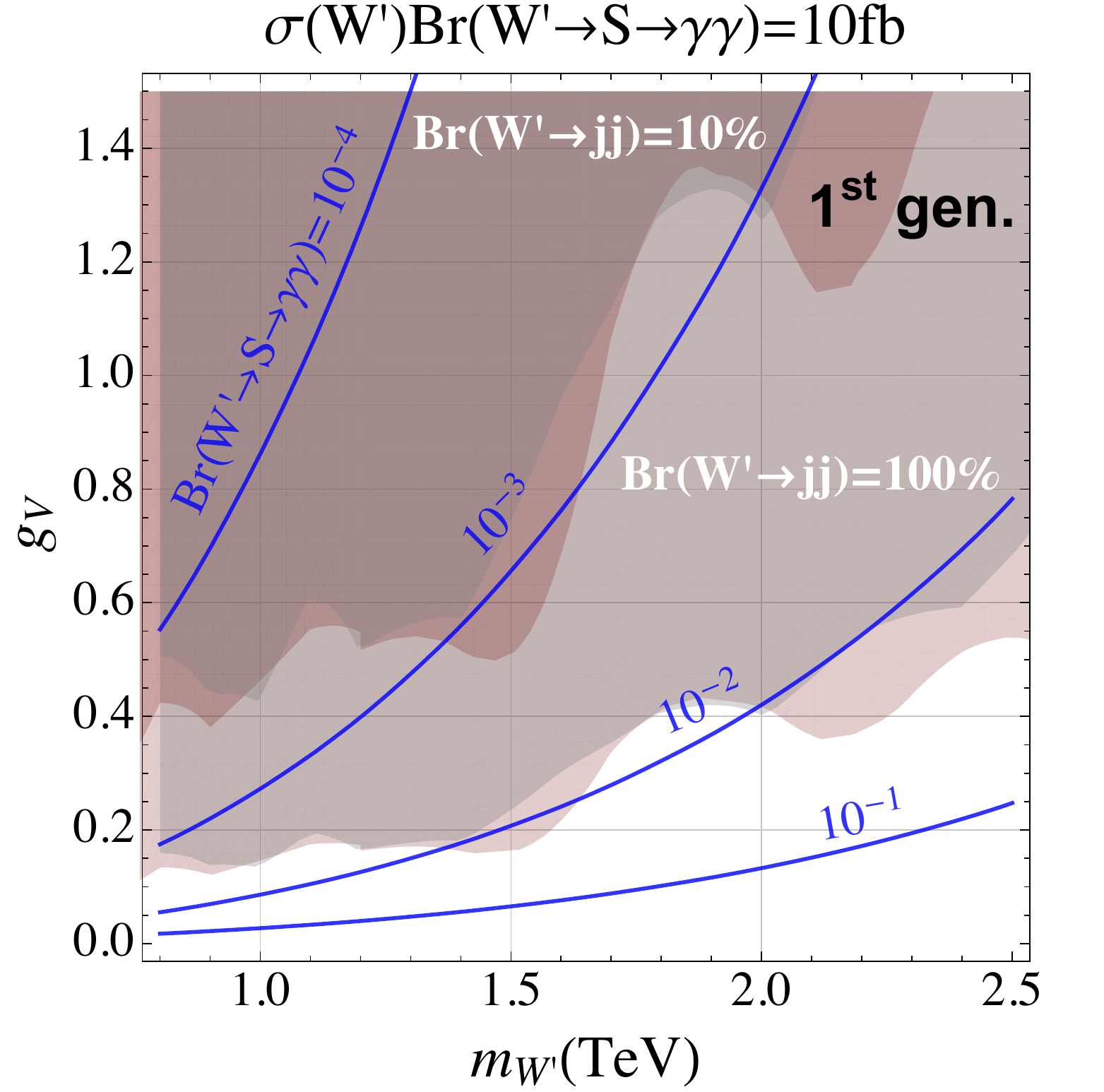} \\[9pt]
\includegraphics[width=0.45\textwidth]{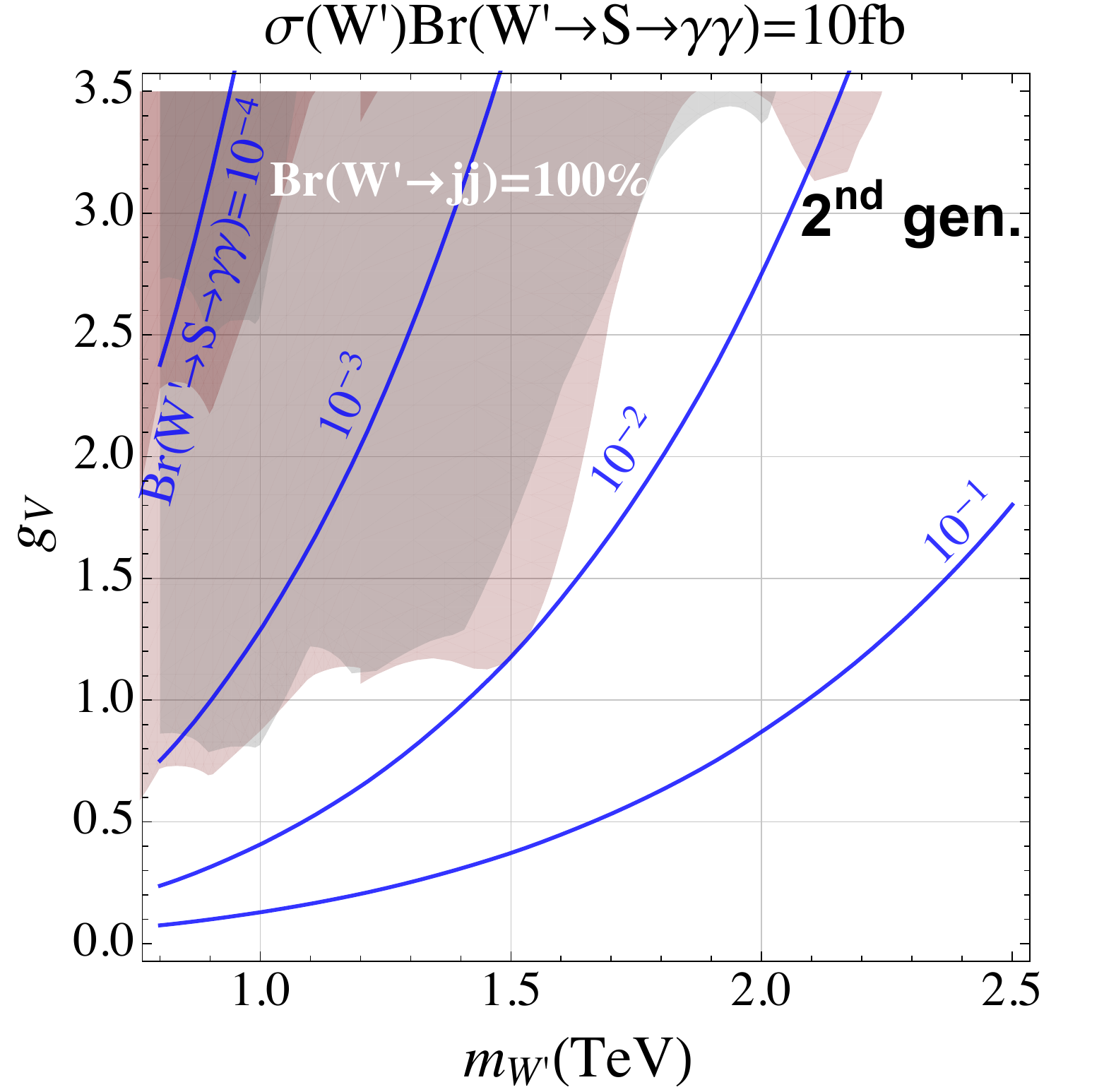}
\caption{Top: The blue solid lines show  $g_V=g_{R,u}^{W'}$ required to obtain the di-photon signal cross section of $10 {\rm fb}$ as a function of $W'$ mass, assuming combined branching ratios $Br(W'\to \Hc^+\Sc(\to \gamma\gamma))=10^{-1}/10^{-2}/10^{-3}/10^{-4}$ (bottom to top), and setting to zero the other couplings of SM fermions to $W'$. The grey (red) shaded regions show di-jet exclusions from ATLAS \cite{Aad:2014aqa} (CMS \cite{Khachatryan:2015sja,CMS-PAS-EXO-14-005}) 8TeV data assuming $Br(W'\to jj)=100\%)$ (10\% for darker regions). Bottom: the same as the top but assuming $W'$ only couples to the 2nd generation right handed quarks.}
\label{fig:gVmW'}
\end{figure}

We turn next to  the Drell-Yan production, which we describe with a phenomenological Lagrangian containing an  $SU(2)_L$ triplet of heavy vector resonances, $\Vc_\mu^i$, a scalar weak triplet, $\Hc^i$, and a scalar weak singlet, $\Sc$. They are assumed to be color and Hypercharge singlets. The diphoton signal will be due to the $pp\to \Vc^i \to \Hc^i \Sc$, with either $\Sc$ or the neutral component of $\Hc_3$ decaying to two photons. In this section we are agnostic about whether the fields $\Hc_\mu^i, \Hc^i, \Sc$ are fundamental or composite. 
As an example of a perturbative model one can consider the $SU(2)_L\times SU(2)_R\times U(1)'$ that were invoked for the explanation of the putative $e^+e^-$, $jj$, $Wh^0$ and $WZ$ anomalies near 1.9 TeV 
in the 8 TeV data \cite{Dobrescu:2015jvn,Dobrescu:2015yba,Brehmer:2015cia,Dev:2015pga}. In this case $\Vc_\mu^i$ are the gauge bosons of $SU(2)_R$, while $\Hc^i$ and $S$ are additional scalars not needed in \cite{Dobrescu:2015jvn,Dobrescu:2015yba,Brehmer:2015cia,Dev:2015pga}. One can also imagine that the $\Vc_\mu^i$ are the lightest vector resonances of the new strongly interacting sector, while $\Hc^i$ and $\Sc$ are the pseudo-Goldstone bosons due to spontaneous breaking of a global symmetry in the composite sector.

In our phenomenological description we only need the transformation properties of $\Vc_\mu^i$, $\Sc$ and $\Hc^i$ under the SM gauge group. The vectors couple to the SM fermions, $f_L$ and $f_R$, through
\beq
{\cal L}_{\rm eff}\supset g_{L,f}^{\Vc^i} \Vc_\mu^i \bar f_L \gamma^\mu \tau^i f_L +g_{R,f}^{\Vc^i} \Vc_\mu^i \bar f_R \gamma^\mu \tau^i f_R.
\eeq
In the notation we distinguish the couplings of $\Vc_\mu^3=Z_\mu'$ from those of $(\Vc_\mu^1\mp i \Vc_\mu^2)/\sqrt 2=W_\mu'{}^\pm$, since they differ slightly due to $U(1)_Y$ breaking of the custodial symmetry. The $Z'$ coupling to SM fermions can come from the $Z'$--$Z$ mixing, e.g., through the operator
\beq\label{eq:g'VHDH}
g_\Vc \Vc_\mu^i H^\dagger \tau^i D_\mu H,
\eeq
where $H$ is the SM Higgs doublet. After EWSB this term introduces a mass mixing between $Z$ and $Z'$, resulting in the mass matrix
\beq
{\cal M}_{ZZ'}=
\begin{pmatrix}
m_Z^2 & g_\Vc v_W m_Z\\
g_\Vc v_W m_Z & m_{Z'}^2
\end{pmatrix}.
\eeq
The couplings of $Z'$ to the SM fermions are thus
\beq
g_{L,f}^{Z'}=g_{L,f}^{Z} \sin\theta_{ZZ'}, \quad g_{R,f}^{Z'}=g_{R,f}^{Z} \sin\theta_{ZZ'},
\eeq
where $g_{L,f}^Z$ and $g_{R,f}^Z$ are the couplings of $Z$ to the left-handed and right-handed SM fermions, and the mixing angle is given by
\beq
\sin \theta_{ZZ'}\simeq \frac{g_\Vc v_W m_Z}{m_{Z'}^2}.
\eeq
The expressions for couplings to $W'$ are obtained from the above by replacing $Z\to W$.

The $Z$--$Z'$ and $W$--$W'$ mixing also receives contributions from the kinetic mixing, 
\beq
{\cal L}\supset \frac{1}{2} \epsilon_{\rm kin} \Vc_{\mu\nu}^i W_{\mu\nu}^i,
\eeq
where $\Vc_{\mu\nu}^i =D_\mu \Vc_\nu^i- D_\nu \Vc_\mu^i$, while $W_{\mu\nu}^i =\partial_\mu W_\nu^i -\partial_\nu W_\mu^i +i g_2 \epsilon^{ijk} W_\mu^i W_\nu^j$ are the components of the $SU(2)_L$ field strength. The kinetic mixing term can be removed by shifting the $W_\mu^i$ fields
\beq
W_\mu^i\to W_\mu^i+\epsilon_{\rm kin} \Vc_\mu^i.
\eeq
After these redefinitions, the $\Vc_\mu^i$ couples to the SM weak currents with the strength $\epsilon_{\rm kin}$. The couplings to $W'$ are given by 
\beq
g_{L,f}^{W'}=g_{L,f}^{W} \epsilon_{\rm kin}, \quad g_{R,f}^{W'}=g_{R,f}^{W} \epsilon_{\rm kin},
\eeq
and similarly for $Z'$, obtained by replacing $W\to Z$ in the above expression. In perturbative theories the kinetic mixing parameter $\epsilon_{\rm k}$ is loop suppressed, $\epsilon_{\rm k}\sim 1/(16\pi)^n$, where $n$ is the number of loops required to connect $W_\mu^i$ and $\Vc_\mu^i$. In strongly interacting theories, the mixing can be large, $\epsilon_{\rm kin}\sim {\mathcal O}(1/4\pi)$. 

The coupling of $\Vc_\mu^i$ to $\Hc^i$ and $\Sc$ is given by
\beq
{\cal L}_{\rm eff}\supset g_{\Vc\Hc\Sc} \Vc_\mu^i D_\mu \Hc^i \Sc+ g_{\Vc\Hc\Hc} \epsilon^{ijk} \Vc_\mu^i D_\mu \Hc^j \Hc^k.
\eeq
The scalar $\Sc$ couples to SM gauge bosons through dimension five operators, given in \eqref{eq:Seffective}. The scalar triplets have the following couplings to the SM gauge bosons
\beq
\begin{split}
{\cal L}_{\rm eff}\supset & \tilde \lambda_{WB} \frac{\alpha}{\pi m_{\Hc^i}}\Hc^i W_{\mu\nu}^i \tilde B_{\mu\nu}\\
&+\tilde \lambda_{WW} \frac{\alpha}{\pi m_{\Hc^i}}\Hc^i W_{\mu\nu}^j \tilde W_{\mu\nu}^k \epsilon_{ijk},
\end{split}
\eeq
where we assumed that $\Hc^i$ are pseudoscalars, as realized in strongly interacting models.

In Fig. \ref{fig:gVmW'} we show the required couplings to $W'$ in order to have a di-photon signal at the rate 
$\sigma(W'{}^\pm)\times Br(W'{}^\pm\to \Hc^\pm \Sc) Br (\Sc\to \gamma\gamma)=10\, {\rm fb}.$
On the top panel, we consider the limit where $W'$ only couples to the right-handed $u$ and $d$ quarks, and 
 thus set
 \beq
g_{R,1}^{W'}=g_V, \qquad g_{R,i\ne 1}^{W'}= g_{L,i}^{W'}=0.
\eeq
The blue solid lines show the value of $g_V$ required to obtain a diphoton signal of $10$ fb as a function of $W'$ mass, if the branching ratio $Br(W'{}^\pm\to \Hc^\pm \Sc) Br (\Sc\to \gamma\gamma)$ is set to $10^{-1}, 10^{-2}, 10^{-3}, 10^{-4}$ (bottom to top in the figure). The shaded regions are excluded by dijet searches, assuming that $W'$ decays almost exclusively through $W' \to jj$ (the dark shaded regions show exclusions for $Br(W'\to jj)=0.1$). The gray shaded region shows exclusion from ATLAS narrow dijet resonance search  \cite{Aad:2014aqa}, while red shaded region shows the CMS exclusions \cite{Khachatryan:2015sja,CMS-PAS-EXO-14-005}, both with 8TeV data. Equivalent plot, but for $W'$ coupling only to right-handed $c$ and $s$ quarks, is shown in the bottom panel of Fig. \ref{fig:gVmW'}. 

If $W'$ predominantly decays to di-jets, a branching ratio $Br(W'{}^\pm\to \Hc^\pm \Sc) Br (\Sc\to \gamma\gamma)\gtrsim 1\%$ is required. For a $W'$ resonance with mass of 2 TeV (1.5 TeV) a coupling to $u$ and $d$ quarks of $g_V\sim 0.4\, (0.2)$ is required (if $W'$ couples vectorially to both left-handed and right-handed quarks these values are smaller by $\sqrt 2$). If $W'$ has other dominant decay channels, the coupling $g_V$ can be correspondingly larger, and by $Br(W'{}^\pm\to \Hc^\pm \Sc) Br (\Sc\to \gamma\gamma)$ can be smaller. For instance, if $Br(W'\to j j)=10\%$, then $Br(W'{}^\pm\to \Hc^\pm \Sc) Br (\Sc\to \gamma\gamma)\sim 10^{-3}$ is still allowed by the dijet searches. Note that to a good approximation Fig. \ref{fig:gVmW'} (top) applies also to the case where $W'$ couples universally to all three generations, due to the suppressions of the $s$, $c$, and $b$ pdfs. The effect of the pdf suppression is clearly visible when the top and bottom pannels of Fig. \ref{fig:gVmW'}. If $W'$ couples only to $c$ and $s$ quarks then $g_V$ needs to be about $\sim 5\times$ larger than if it couples to $u$ and $d$ quarks.

\section{Conclusions} \label{sec:conclusions}

The di-photon excess can be explained by a scalar singlet coupled to gluons and electroweak vector bosons through effective dimension 5 operators. We showed that a large enough cross section and consequently the diphoton signal is obtained if either (i) the scalar is produced through gluon fusion, or even (ii) if the singlet is produced entirely through vector boson fusion. One possibility to induce the dimension 5 couplings is through loops of vector-like fermions. Depending on the electroweak gauge quantum numbers of the fermions, one also expects signals in $WW$, $ZZ$ and $Z\gamma$ resonance searches with cross sections that are comparable to the observed $\gamma\gamma$ signal. Some scenarios are already constrained by 8~TeV resonance searches in the $WW$, $ZZ$ and $Z\gamma$ final states. The prospects for detecting resonances in these channels at the 13~TeV run of the LHC are in general excellent.
If the couplings of the vector-like fermions to the singlet are of $\Ord(1)$ and their charges are not exotically large, several copies of vector-like fermions are required to induce large enough effective gluon and photon couplings. The effective couplings decouple as $v_W/m_f$, where $m_f$ is the mass of the vector-like fermions. We expect these fermions not to be far above the TeV scale and potentially within direct reach of the LHC at 13 TeV. 

In the context of two Higgs doublet models, both the heavy scalar $H$ and the heavy pseudoscalar $A$ can in principle produce a diphoton signal. If the second Higgs doublet is the only new degree of freedom beyond the Standard Model, we find that the signal cross sections are typically orders of magnitude below the observed excess. Adding charged and colored degrees of freedom (e.g. in the form of vector-like fermions) allows for large enough gluon and photon couplings to the second doublet in order to explain the data. However, in the doublet case the new physics contributions to the effective gluon and photon couplings decouple as $v_W^2/m_f^2$. Therefore a very large number of additional degrees of freedom is required in order to induce large enough couplings, rendering an explanation in the context of a 2HDM less plausible.

An alternative possibility for the diphoton excess is that it is due to a cascade decay,  $pp\to X\to Y(\to\gamma\gamma) Y'$. The heavier resonance, $X$, can either be produced through gluon fusion, or through Drell-Yan production. The searches for dijet resonances at 8TeV place strong constraints on the allowed parameter space of the models. If $X$ decays predominantly to dijets, then generically $Br(Y\to\gamma\gamma)$ needs to be above $10^{-2}$. This may be a challenge in models that address naturalness, but can be avoided in ad-hoc models invoked to explain the di-photon excess.

\mysection{Acknowledgements}
The work of A.K. is supported by the DOE grant DE-SC0011784. J.Z. is supported in part by the U.S. National Science Foundation under CAREER Grant PHY-1151392. The research of W.A. and S.G. at Perimeter Institute is supported by the Government of Canada through Industry Canada and by the Province of Ontario through the Ministry of Economic Development \& Innovation. The work of A.M. was partially supported by the National Science Foundation under Grant No. PHY-1417118. J.G. is supported by the James Arthur Postdoctoral Fellowship at NYU.

 \bibliography{paper_ref}

\end{document}